\documentclass[]{aa}
\usepackage[dvips]{graphicx}
\usepackage{txfonts}
\begin{document}
\title{Tests of stellar model atmospheres by optical interferometry III}
\subtitle{NPOI and VINCI interferometry of the M0 giant $\gamma$\,Sge covering
0.5 - 2.2\,$\mu$m
\thanks{Based on data obtained with the Navy Prototype Optical Interferometer
(NPOI). The NPOI is a joint project of the Naval Research Laboratory and 
the United States Naval Observatory in cooperation with Lowell Observatory, 
and is funded by the Office of Naval Research and the Oceanographer of 
the Navy.}\fnmsep
\thanks{Based on public commissioning data released by 
the Paranal Observatory, Chile, and received via the
ESO/ST-ECF Science Archive Facility.}
}
\author{
M.~Wittkowski\inst{1} \and
C.~A.~Hummel\inst{2} \and
J.~P.~Aufdenberg\inst{3} \and
V.~Roccatagliata\inst{4}
}
\institute{
European Southern Observatory, Karl-Schwarzschild-Str. 2,
85748 Garching bei M\"unchen, Germany, \email{mwittkow@eso.org}
\and
European Southern Observatory, Casilla 19001, Santiago 19, Chile
\and
National Optical Astronomy Observatory, 950 North Cherry Avenue,
Tucson, AZ 85719, USA
\and
Max-Planck-Institut f\"ur Astronomie, K\"onigsstuhl 17, 
69117 Heidelberg, Germany
}
\titlerunning{NPOI \& VLTI Interferometry of $\gamma$\,Sagittae}
\date{Received \dots; accepted \dots}
\abstract{Optical interferometry allows a measurement of the 
intensity profile across a stellar disc, leading to a direct test 
and calibration of theoretical model atmospheres as well as to a 
precise determination of fundamental stellar parameters.}
{We present a comparison of the visual and near-infrared intensity profile 
of the M0 giant $\gamma$\,Sagittae to plane-parallel \protect{\tt ATLAS\,9} 
as well as to plane-parallel \& 
spherical \protect{\tt PHOENIX} model atmospheres.}
{We use previously described visual interferometric data obtained 
with the Navy Prototype Optical Interferometer (NPOI) in July 2000.
We apply the recently developed technique of \protect{\it coherent integration}, 
and thereby obtain visibility data of more spectral channels 
(526-852\,nm) and with higher precision than before. 
In addition, we employ new measurements of the 
near-infrared $K$-band ($\sim$\,2200\,nm) diameter
of $\gamma$\,Sagittae obtained with the instrument VINCI at the 
ESO VLT Interferometer (VLTI) in 2002.}
{The spherical \protect{\tt PHOENIX} model leads to a precise definition
of the Rosseland angular diameter and a consistent 
high-precision diameter value for our NPOI and VLTI/VINCI data sets of
$\Theta_\mathrm{Ross}=6.06\ \pm\ 0.02$\,mas, with the Hipparcos parallax
corresponding to $R_\mathrm{Ross}=55\ \pm\ 4\ R_\odot$, and with the 
bolometric flux corresponding to an effective temperature 
$T_\mathrm{eff}=3805\ \pm\ 55$\,K.
Our visual visibility data close to the first minimum and in the 
second lobe constrain the limb-darkening
effect and are generally consistent with the model atmosphere predictions. 
The visual closure phases exhibit a smooth transition between 0 and $\pi$.}
{The agreement between the NPOI and VINCI diameter values increases 
the confidence in the model atmosphere predictions from optical to 
near-infrared wavelengths as well as in the calibration and accuracy
of both interferometric facilities. The consistent night-by-night diameter
values of VINCI give additional confidence in the given uncertainties.
The closure phases suggest a slight deviation from circular symmetry, which 
may be due to surface features, an asymmetric extended layer, or a 
faint unknown companion.}
\keywords{Techniques: interferometric -- Stars: late-type -- 
Stars: AGB and post-AGB -- Stars: atmospheres -- Stars: fundamental 
parameters -- Stars: individual: \object{$\gamma$ Sagittae}}
\maketitle
\section{Introduction}
\label{sec:introduction}
Cool giants on the red giant branch (RGB) and asymptotic
giant branch (AGB) are very luminous and
extended, have a low surface temperature, and their atmospheres can 
thus be rich in molecules. Cool giants are the most important 
source of dust formation and its delivery to the interstellar medium. 
The detailed structure of their extended atmospheres, including the  
effects from circumstellar molecular and dust layers, are still a matter of 
investigation and debate (cf., e.g. Scholz \cite{scholz85,scholz98,scholz01},
Perrin et al. \cite{perrin04}, Ohnaka \cite{ohnaka04a}, 
Ireland \& Scholz \cite{ireland06}).

Theoretical atmosphere models predict in general the spectrum emerging 
from every point of a stellar disc.
Optical interferometry provides the strongest observational
constraint of this prediction by resolving the stellar disc. 
In addition, the constraints on the intensity profiles allow us to find
meaningful definitions of the stellar radius and its precise measurement. 

For regular cool non-pulsating giants, the centre-to-limb variation (CLV) is 
mainly characterised by the limb-darkening effect, which is an effect 
of the vertical temperature profile of the stellar atmosphere. The 
strength of the limb-darkening can be probed by 
optical interferometry in two ways (cf., e.g. 
Hanbury Brown et al. \cite{hanburybrown74}, 
Quirrenbach et al. \cite{quirrenbach96}, Burns et al. \cite{burns97},
Hajian et al. \cite{hajian98}, Wittkowski 
et al. \cite{wittkowski01,wittkowski04}, 
Aufdenberg et al. \cite{aufdenberg05}):
(1) by measuring variations of an equivalent uniform disc diameter
(i.e. the uniform disc that has the same integral flux as the true
intensity profile) as a function of wavelength, and
(2) by directly constraining the star's intensity profile in the second 
and higher lobes of the visibility function at one or several bandpasses. 

It was found that pulsating giants as well as supergiants may
exhibit more complex intensity profiles at near- and mid-infrared wavelengths,
showing Gaussian-shaped intensity profiles, tail-like extensions to a
photospheric intensity profile, and multiple components, such as 
a photosphere plus a circumstellar 
shell (cf., e.g. Woodruff et al. \cite{woodruff04}, 
Ohnaka \cite{ohnaka04a,ohnaka04b},
Perrin et al. \cite{perrin04,perrin05}, Fedele et al. \cite{fedele05}).
Additionally, observed intensity profiles might be affected by
dust shells (e.g. Ohnaka et al. \cite{ohnaka05}, 
Ireland \& Scholz \cite{ireland06}) or horizontal surface inhomogeneities
(e.g. Burns et al. \cite{burns97}).

In Wittkowski et al. (\cite{wittkowski01}, hereafter Paper\,I), we
used the Navy Prototype Optical Interferometer (NPOI, 
Armstrong et al. \cite{armstrong98}), used the method of baseline 
bootstrapping (cf. Hajian et al. \cite{hajian98}), and developed 
improved methods of compensation of noise and detection bias terms, 
in order to obtain precise visual visibility measurements in the second lobe 
of the visibility function for three cool giants. 
We found agreement with predictions by plane-parallel 
ATLAS\,9 (\cite{kurucz93}) model atmospheres within the obtained
wavelength range and precision.
Thereby, the strength of the  
limb-darkening effect and the stars' fundamental parameters were constrained.
Aufdenberg \& Hauschildt (\cite{aufdenberg03})  compared one of the NPOI 
observations of $\gamma$\,Sagittae from Paper\,I
to a spherical {\tt PHOENIX} (\cite{hauschildt99}) model atmosphere
and found agreement.
In Wittkowski et al. (\cite{wittkowski04}, hereafter Paper II), we
directly measured the limb-darkening effect of the M4 giant $\psi$\,Phoenicis
using the ESO Very Large Telescope Interferometer (VLTI) in the near-infrared
$K$-band, confronted the observations with predictions by independently 
constructed {\tt ATLAS\,9} and {\tt PHOENIX} model atmospheres, and found
agreement with all considered models.

Recently, Hummel et al. (\cite{hummel03}) developed the method of 
{\it coherent integration} and its application to NPOI data in order to increase the 
precision of visibility measurements. This method was recently applied 
by Peterson et al. (\cite{peterson06a,peterson06b}) to NPOI observations 
of Altair and Vega.

Here, we reanalyse the NPOI data of the M0 
giant $\gamma$\,Sagittae (HR\,7635, HD\,189319), the 
brightest of the targets in Paper~I, using the newly developed method 
of {\it coherent integration}.
We obtain visibility data with higher precision than in Paper~I, and --due 
to the lower noise-- are able to make use of more spectral channels toward
the blue end of NPOI's wavelength range. Now, the wider wavelength range
covers 526-852\,nm, compared to 649-852\,nm in Paper I.
We thus also increase our maximum spatial resolution from 
$\approx$3.3\,mas to $\approx$2.7\,mas, which gives important additional
visibility data in the second lobe that are sensitive to the limb-darkening
effect. 
In addition, we observed  $\gamma$\,Sagittae with the ESO Very Large
Telescope Interferometer (VLTI) and its $K$-band instrument VINCI, in order
to compare results derived from different interferometric facilities, and
to probe the consistency of the wavelength-independent Rosseland diameter
from visual to near-infrared wavelengths. 

The cool giant $\gamma$\,Sagittae does not appear
in the Combined General Catalogue of Variable 
Stars (Samus et al. \cite{samus04}), indicating that it lacks strong
photometric variability. Thus, it is a good target for the purpose of 
calibrating model atmospheres and deriving high-precision fundamental
parameters. The spectral type has been listed as K5-M0\,III by
Morgan \& Keenan (\cite{morgan73}), and been revised to M0\,III by
Keenan \& McNeil (\cite{keenan89}).
Wisniewski \& Morrison (private communication) confirm by means of optical
echelle spectra recently obtained at Ritter Observatory 
that $\gamma$\,Sagittae's
spectrum closely resembles that of the MK standard $\mu$\,UMa (M0\,III).
We determine the bolometric flux of $\gamma$\,Sagittae
to $f_\mathrm{bol}=(2.57\pm 0.13)\times 10^{-9}$\,W/m$^2$ by means of
a spline fit and integration of the narrow-band spectrophotometric data
by Alekseeva et al. (\cite{alekseeva97}) covering 405\,nm to 1080\,nm
complemented by broadband photometry shortward and longward of 
this range from the 13-colour photometry by Johnson et al. (\cite{johnson75}).
The values for $f_\mathrm{bol}$ of $(2.79 \pm 0.14)\times 10^{-9}$\,W/m$^2$
and $(2.83 \pm 0.14)\times 10^{-9}$\,W/m$^2$ by Alonso et al. (\cite{alonso99})
and Mozurkewich et al. (\cite{mozurkewich03}), respectively, are derived 
from broad-band photometry alone and likely overestimate $f_\mathrm{bol}$ 
because of a too sparse sampling of the visual spectrum including the TiO 
band heads and other features.
The limb-darkened angular diameter of $\gamma$\,Sagittae has been determined 
in Paper\,I to be 6.18\,$\pm$0.07\,mas, based on a comparison of NPOI 
visibility data to {\tt ATLAS\,9} model atmospheres. This value corresponds 
to a limb-darkened radius of 56\,$\pm$\,4\,$R_\odot$, derived with the 
Hipparcos parallax of 11.90\,$\pm$\,0.71\,mas 
(Perryman \& ESA, \cite{perryman}). 
These values of angular diameter, absolute radius, and bolometric flux 
constrain the effective temperature 
to $T_\mathrm{eff}=$ 3768\,K $\pm$ 70\,K, and the
luminosity to $\log L/L_\odot=$ 2.75 $\pm$ 0.10. Placing $\gamma$\,Sagittae
on the Hertzsprung Russel diagram using these values, and comparing to
stellar evolutionary tracks by Girardi et al. (\cite{girardi00}) as
in Paper\,II (Fig.~1 of Paper\,II) we can estimate a mass 
of $M=$ 1.3 $\pm$ 0.4 $M_\odot$, and thus a surface gravity 
of $\log g$=1.06 $\pm$ 0.22. These values are used as an a-priori estimate
for our analysis and will be refined in the conclusions. 
\section{NPOI measurements}
\subsection{NPOI observations}
We reanalyse the visual $\gamma$ Sagittae data in Paper\,I obtained 
with NPOI on July 21, 2000.
The centre (C), east (E) and west (W) siderostats of
the astrometric sub-array of NPOI were used to obtain baselines
of ground length 18.9 m (CE), 22.2 m (CW), and 37.5 m (EW).
The data were recorded in 32 spectral channels of equal width
in wavenumber and covering the band from $\approx$ 450\,nm to 850\,nm.
Due to low photon count rates only the 10 reddest channels could be
used in Paper\,I (covering 649\,nm to 852\,nm). 
We reanalyse the same raw data using the newly developed
coherent integration algorithm as first described by 
Hummel et al. (\cite{hummel03}). 
The details of our new analysis are described below.  
The benefits of the new analysis include an improved 
signal-to-noise-ratio (SNR) of the visibility data on the long EW baseline, 
as well as a much improved SNR of the triple amplitudes and phases.
These improvements enable us to use the 20 reddest channels in the present
paper, now covering 526\,nm to 852\,nm.
\subsection{NPOI data reduction and calibration}
\begin{figure}
\centering
\resizebox{\hsize}{!}{\includegraphics{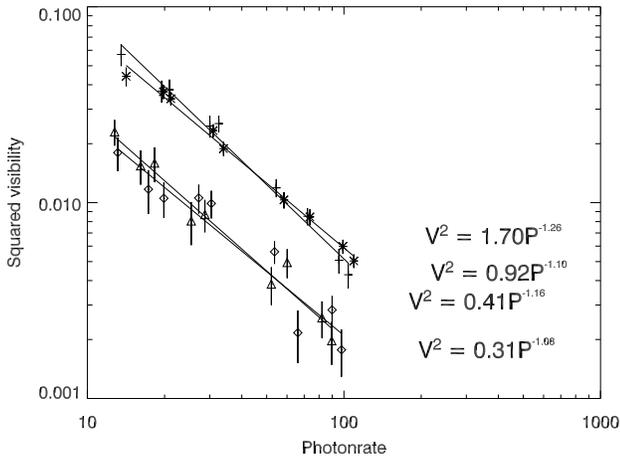}}
\caption{Incoherent integration. 
Off-fringe squared visibility amplitude, i.e. the visibility
bias that remains for data off the fringe packet and that is
compensated after the $Z^2$ compensation (see text for more details).
This bias can be described by a power law as a function of photon rate $P$.
As an example, the residual bias is shown for the four reddest channels
on the EW baseline. Data are 2 ms incoherent integrations from July 22.
Channels 1 though 4 use symbols plus, star, diamond, and triangle, 
respectively. Power-law fit coefficients are given for each channel, 
in the same order as shown on the corresponding plot for the coherent 
analysis in Fig.~\protect\ref{fig:bias200}.}
\label{fig:bias2}
\end{figure}
\begin{figure}
\centering
\resizebox{\hsize}{!}{\includegraphics{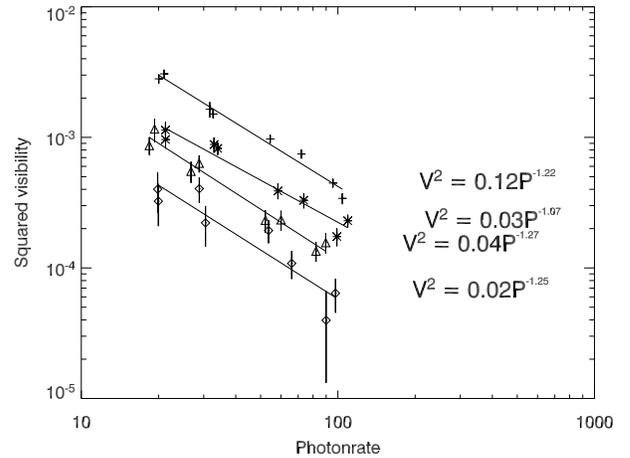}}
\caption{Coherent integration. As Fig.~\protect\ref{fig:bias2}, 
but for 200ms coherent integrations as used in this paper. 
For reasons of comparison, the bin counts are renormalised 
to 2\,ms intervals. It is an additional benefit
of the coherent integration that this residual bias shown here
is clearly reduced compared to the incoherent integration
in Fig.~\protect\ref{fig:bias2}.}
\label{fig:bias200}
\end{figure}

\paragraph{Coherent integration}
The NPOI detector configuration used for our data set (July 2000)
sampled a single fringe of each two telescope interference pattern
using 8 bins every 2 ms. This time interval is called the instrumental
coherent (because it is phase preserving) integration time. Increasing 
this time would eventually lead to a complete loss of fringe contrast 
due to atmospheric fringe motion which is not perfectly compensated by an 
interferometer such as NPOI using the group delay method for fringe tracking.

Therefore, in the so-called incoherent analysis, as used in Paper\,I,
the bin counts of each 2\,ms sample would be Fourier transformed, and an
unbiased estimate for the squared visibility amplitude derived as
follows (see Shao et al. \cite{shao88}):
\begin{equation}
\label{eq:shao}
|V|^2\sim \frac{<X^2+Y^2-\sigma_I^2(N)>}{<N>^2},
\end{equation}
where $X$ and $Y$ are the real and imaginary parts of the Fourier
transform, respectively, and $\sigma_I^2$ is the variance of the
intensity caused by photon and detection noise. In the case of
pure Poisson noise $\sigma_I^2$ equals $N$, the total number of 
bin counts of each sample.
In the case of NPOI, the data of which show non-Poisson noise due to 
afterpulsing of the APDs (avalanche photo diodes), the bias is 
estimated according to procedures described in Paper\,I (see also below
the paragraph ``Correction of noise and detection bias terms'').

The signal to noise ratio SNR of the squared visibility estimator is as 
follows (Shao et al. \cite{shao88}):
\begin{equation}
SNR(V^2)=\frac{1}{4} M^{1/2}NV^2 \left[1+\frac{1}{2}NV^2\right]^{-1/2},
\end{equation}
where $M$ is the number of samples averaged. One can see that by
increasing the coherent integration time, i.e. increasing $N$, 
rather than increasing $M$
a larger gain in SNR can be realised. This is true as long as $NV^2$
is much smaller than unity, otherwise nothing can be gained by a
coherent average over an incoherent average. Here, we chose a coherent
integration time of 200\,ms which still results in $NV^2>5$ for
photon rates of 10 per 2 ms at squared visibilities of about 0.005.
Subsequently, the squared visibility was computed for the coherent
samples, and then incoherently averaged in 2 s intervals. The
complex triple products were computed from the coherent samples
as well, but vector averaged to preserve the phase.
Coherent integration for time intervals longer than the instrumental
coherent integration time (we chose 200 ms)
require the alignment of the complex
visibilities, equivalent to removing the relative offsets between
successive samples of the fringe. As described by 
Hummel et al. (\cite{hummel03}), this can be done in the off-line data 
analysis by making use of the
visibility phase derivative with wavenumber. This quantity, the
so-called group delay, is zero for the white light fringe
which is located at zero relative optical path length difference where
the fringes of all colours interfere constructively.

For the observations described here on $\gamma$ Sagittae, the
importance of coherent integration follows from the very small
visibility amplitudes measured on the long EW baseline since it
samples the second lobe of the Fourier transform of the stellar
disc brightness profile. This baseline is therefore most sensitive
to stellar limb darkening, the focus of this work.
While the low visibility amplitudes on this baseline would prevent
a precise determination of the group delay, the following paragraph
describes how to obtain this estimate in a different way.

\paragraph{Phase bootstrapping}
A design feature of the NPOI array (Armstrong et al. \cite{armstrong98})
is the ability to realise configurations which allow long baselines to 
be boot-strapped by shorter baselines. By this we mean the ability to 
track and observe fringes on long baselines even though the fringe 
contrast can be too low to allow detection of the fringes for 
tracking purposes.
This is achieved by detecting and tracking fringes on the shorter
baselines, and making use of the fact that the sum of all fringe delays
over baselines in a closed loop must equal zero.

In the case described here, the EW baseline is boot-strapped
by the CE and CW baselines. Therefore, the fringe delay on the EW
baseline is equal to the difference of the delay between the
other two baselines, and can thus be computed in this way without
using the measurement on the EW baseline itself.

\paragraph{Correction of noise and detection bias terms}
As shown in Paper\,I, the standard correction for non-Poisson statistics 
of the NPOI detectors used by Hummel et al. (\cite{hummel98})
(for $\sigma_I^2$ of Eq.~\ref{eq:shao}, 
use $Z^2$ with $Z=B_1-B_2+B_3-B_4+B_5-B_6+B_7-B_8$,
where $B_i$ are the bin counts)
can be improved by compensating a small residual bias that remains after
the $Z^2$ compensation. This residual bias is calibrated by a power law
as a function of counts by observing off the fringe for a number of
stars of different brightness (cf. Eqs. 3 and 4 of Paper\,I).
This additional bias correction is important because of the small visibility 
amplitudes measured on resolved stars, where a constant residual bias 
would strongly contribute to the visibility values.
A welcome side effect of the coherent integration is that, because the 
number of photons counted in a coherent integration, $N$, increases, the ratio
of the bias to the signal decreases. Therefore, at a count
rate high enough, the remaining bias after the $Z^2$ compensation
becomes negligible.
Since we chose a coherent integration time of 200\,ms, a measurable,
though much smaller, residual bias than in Paper\,I remained. This 
residual bias was
removed using exactly the same procedures as described in Sect.~3 of 
Paper\,I by reducing data obtained on July 22 during which off-fringe 
data were recorded. Figures~\ref{fig:bias2} and \ref{fig:bias200} show as 
an example for the 4 reddest channels of the EW baseline 
the remaining bias after the $Z^2$ compensation, obtained
for the incoherent analysis as used in Paper\,I (2\,ms integrations)
and for the coherent analysis used here (200\,ms coherent integrations,
and normalised to a 2\,ms interval), respectively.
It can be seen that the residual bias for the coherent analysis 
is clearly reduced compared to the incoherent average. This is an additional
benefit of the method of coherent integration.
\paragraph{Calibration}
The calibration of the data followed the same procedures
as used in Paper\,I. The B6 giant \object{$\epsilon$\,Delphini}, 12 degrees 
away and observed in an interleaved way during the same night, is used as 
the main calibration star. The diameter of $\epsilon$\,Delphini is estimated 
to be 0.3\,mas based on a calibration of the visual magnitude and
$(R-I)$ colour index by Mozurkewich et al. (\cite{mozurkewich91}).
As secondary calibrators we used \object{$\iota^2$\,Cygni} 
and \object{$\pi^2$\,Pegasi}, 
both 32 degrees away from $\gamma$\,Sagittae. The calibration errors
of the squared visibility amplitudes (derived from the scatter
of the primary and secondary calibrator amplitudes and added in 
quadrature to the formal, i.e. photon noise induced errors) are 
now larger than those quoted in Paper\,I for the boot-strapped
EW baseline, namely 13\% versus 7\%, while the calibration error
for the other two baselines was only marginally larger. This is
acceptable since the statistical errors dominate the total error budget 
for the EW baseline.
\begin{figure*}
\centering
\resizebox{0.32\hsize}{!}{\includegraphics{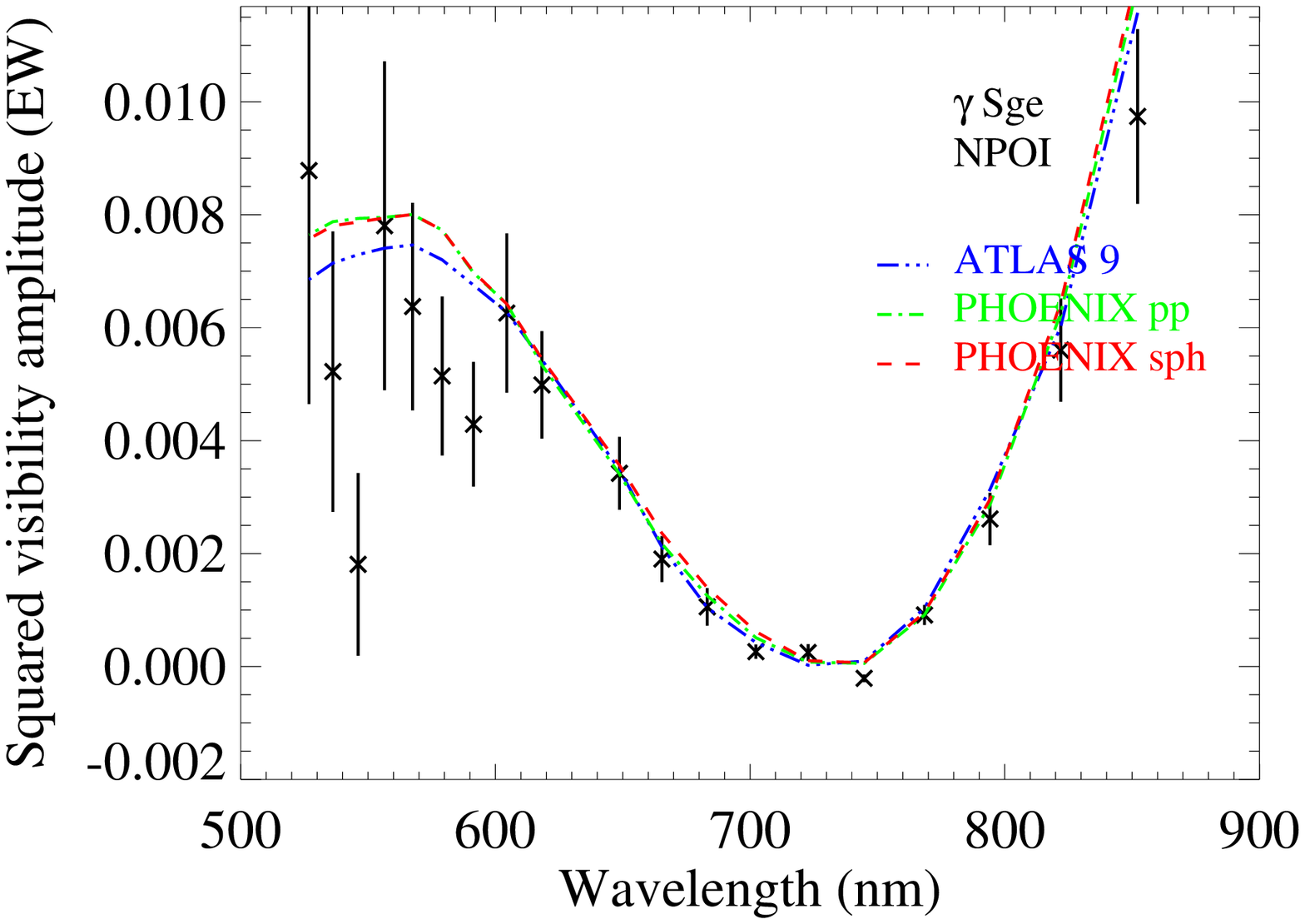}}
\resizebox{0.32\hsize}{!}{\includegraphics{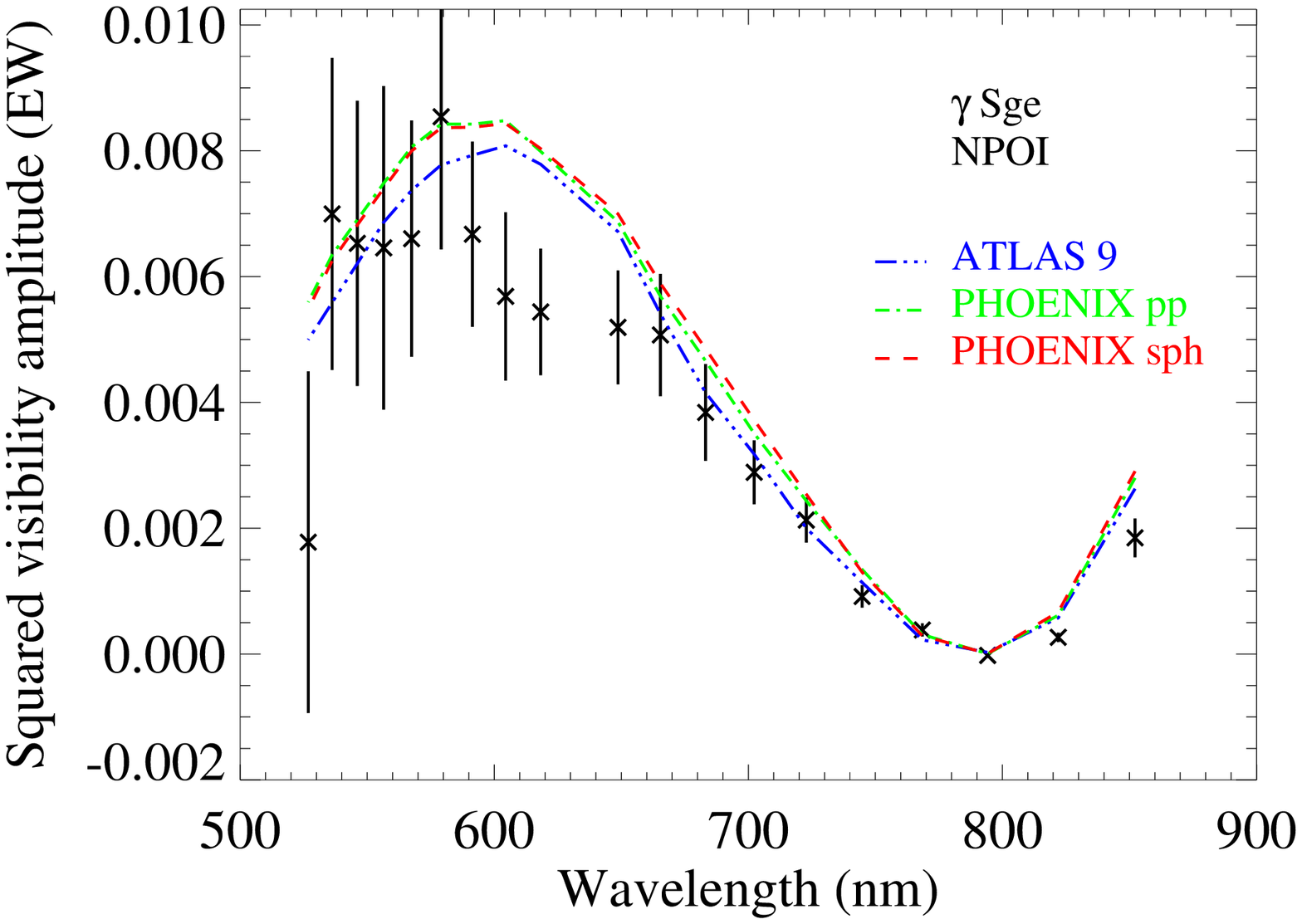}}
\resizebox{0.32\hsize}{!}{\includegraphics{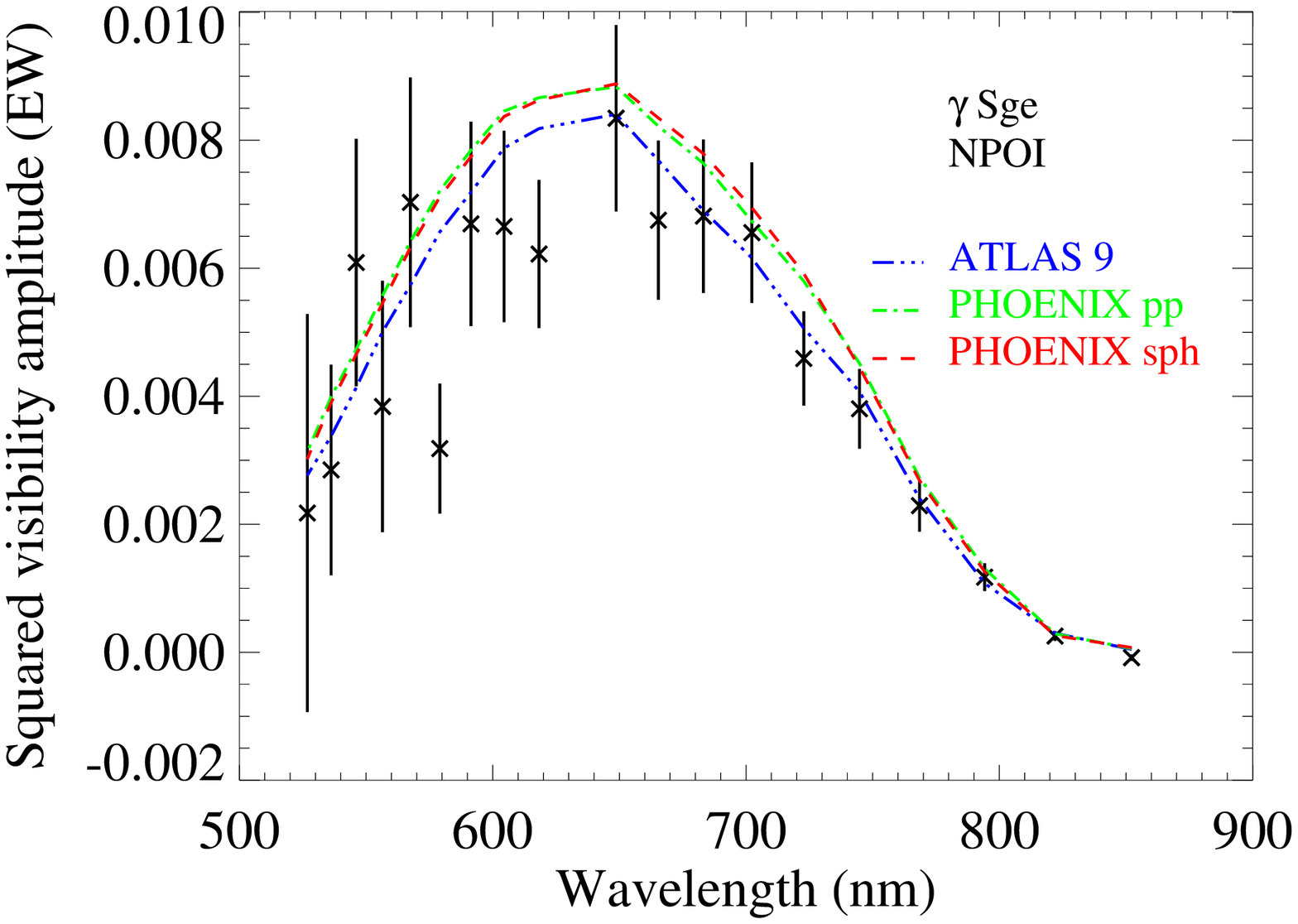}}

\resizebox{0.32\hsize}{!}{\includegraphics{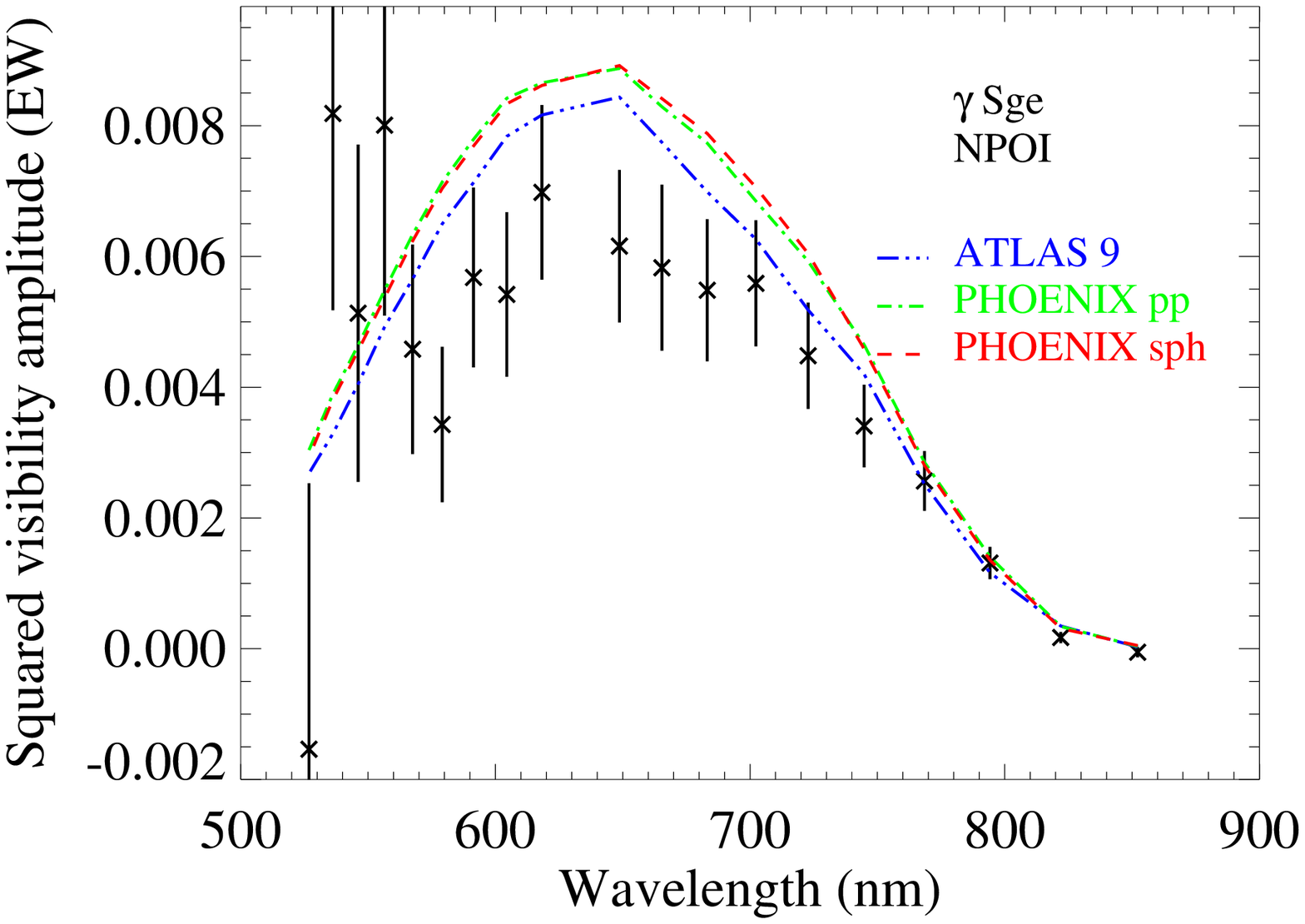}}
\resizebox{0.32\hsize}{!}{\includegraphics{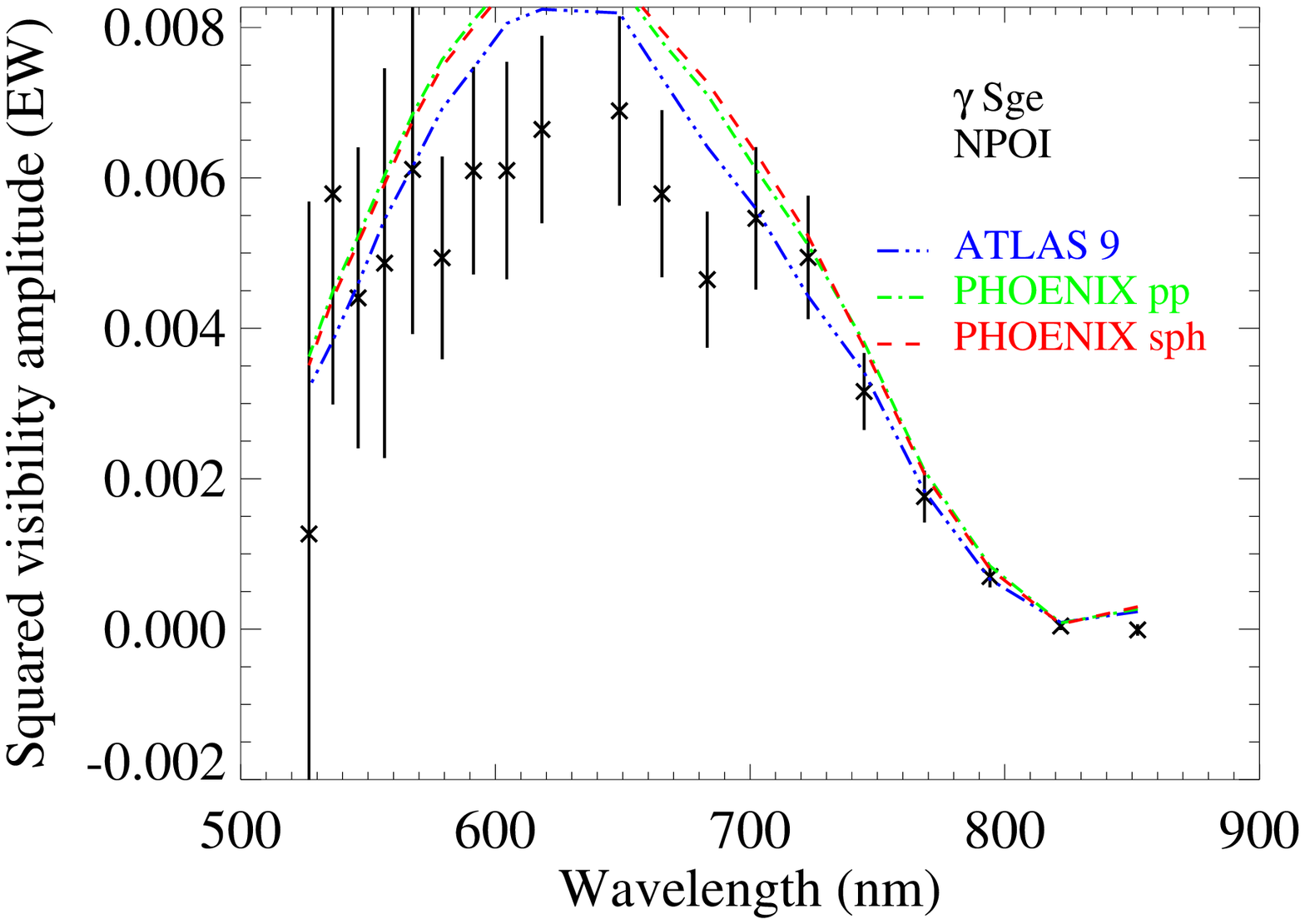}}
\resizebox{0.32\hsize}{!}{\includegraphics{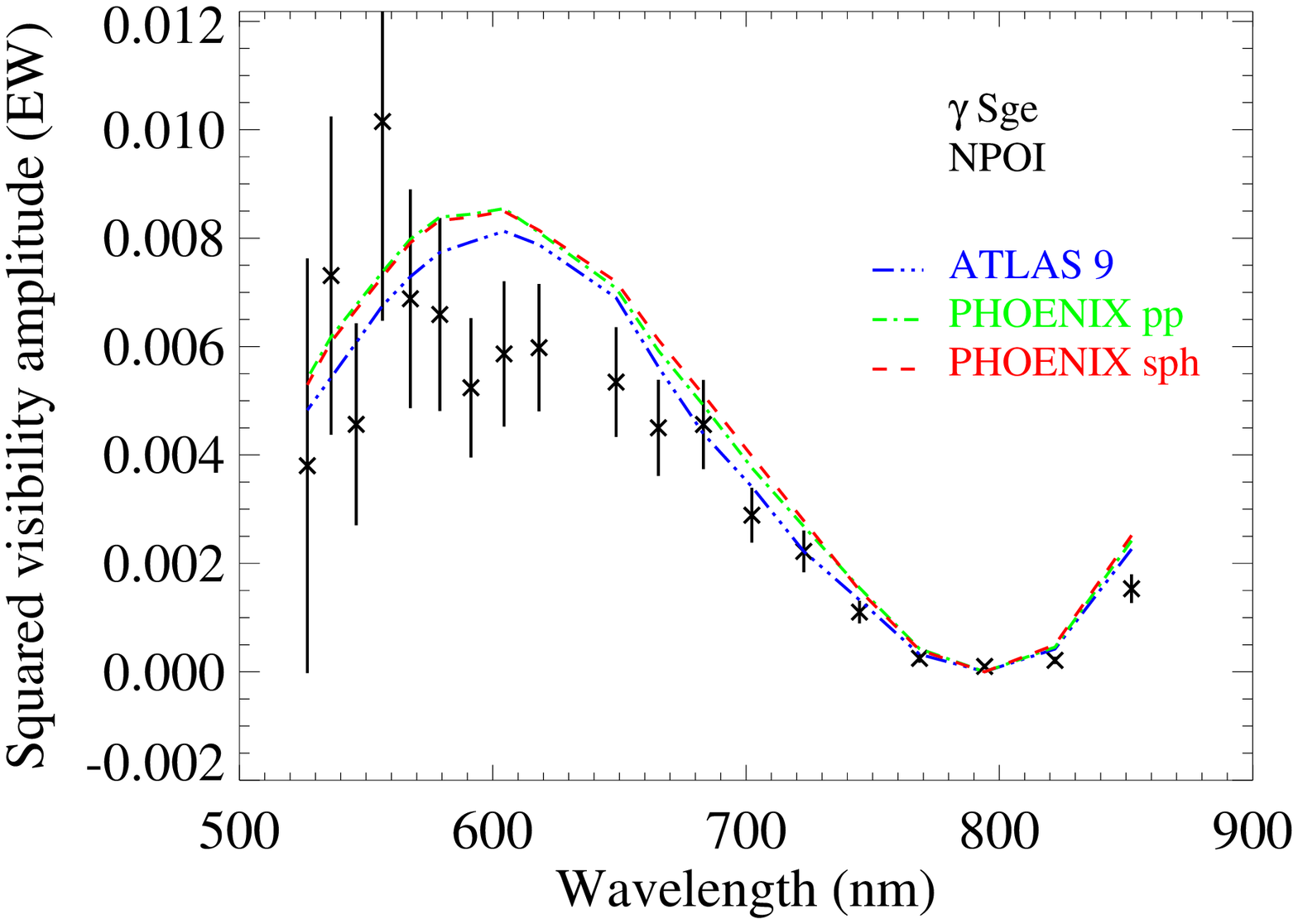}}
\caption{Squared visibility amplitudes of $\gamma$\,Sge obtained from
NPOI on the EW baseline. Also shown are synthetic visibility curves of the 
best fitting model atmospheres as described below 
in Sect.~\protect\ref{sec:models}. 
Each model fit is performed to all NPOI visibility data (squared
visibility amplitudes, triple amplitudes, and closure phases) simultaneously.
The parameters of the plotted model curves are listed 
in Table~\ref{tab:results} .
The synthetic
visibility values are calculated for each specific bandpass of our
observation and these points are connected by straight lines.
A 7th used NPOI observation is not included for the sake of clarity of 
the figure and because its projected baseline length is much shorter and this 
observation thus contains little information in the 2nd lobe.}
\label{fig:npoivis1}
\end{figure*}
\begin{figure*}
\centering
\resizebox{0.32\hsize}{!}{\includegraphics{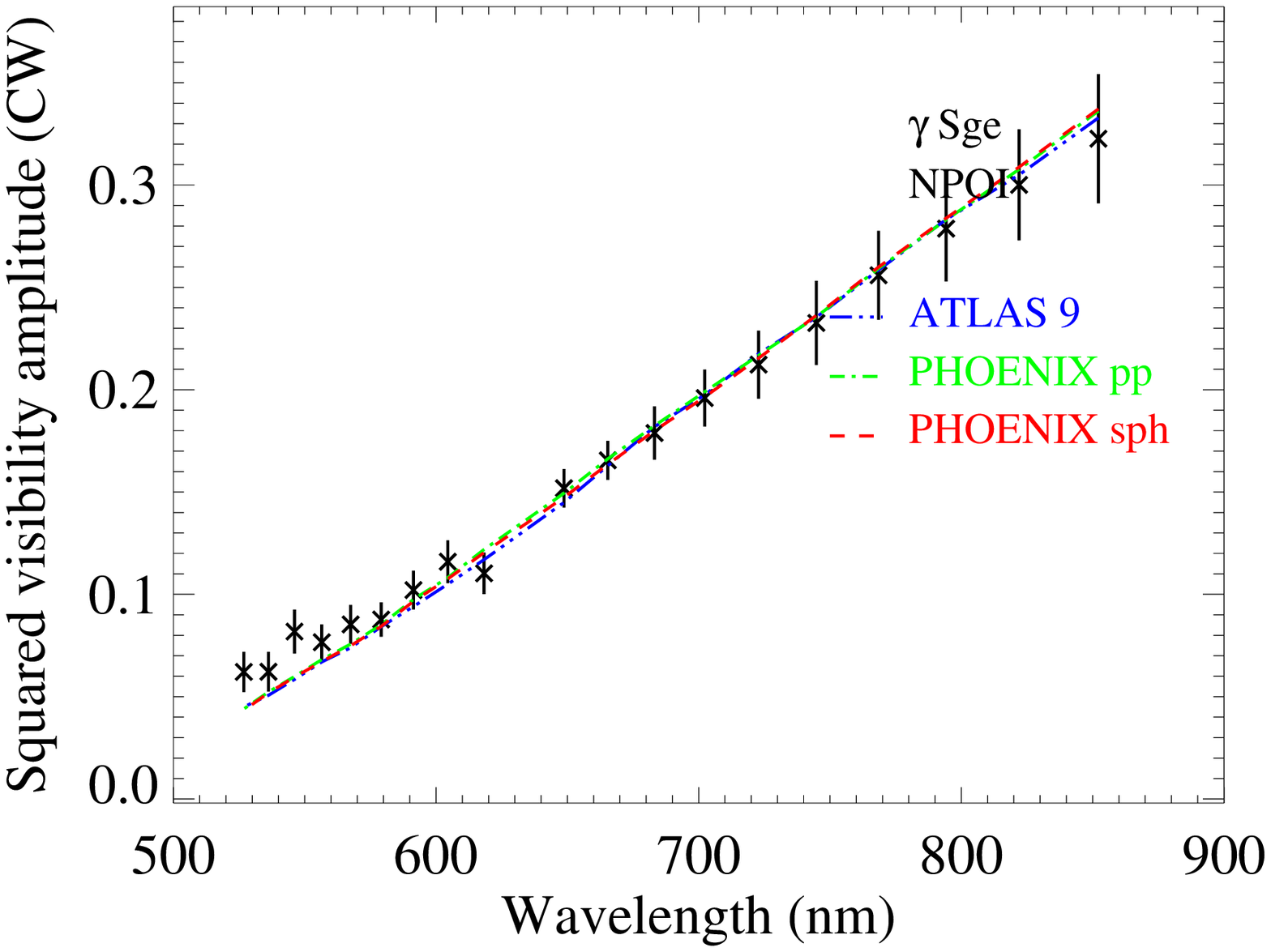}}
\resizebox{0.32\hsize}{!}{\includegraphics{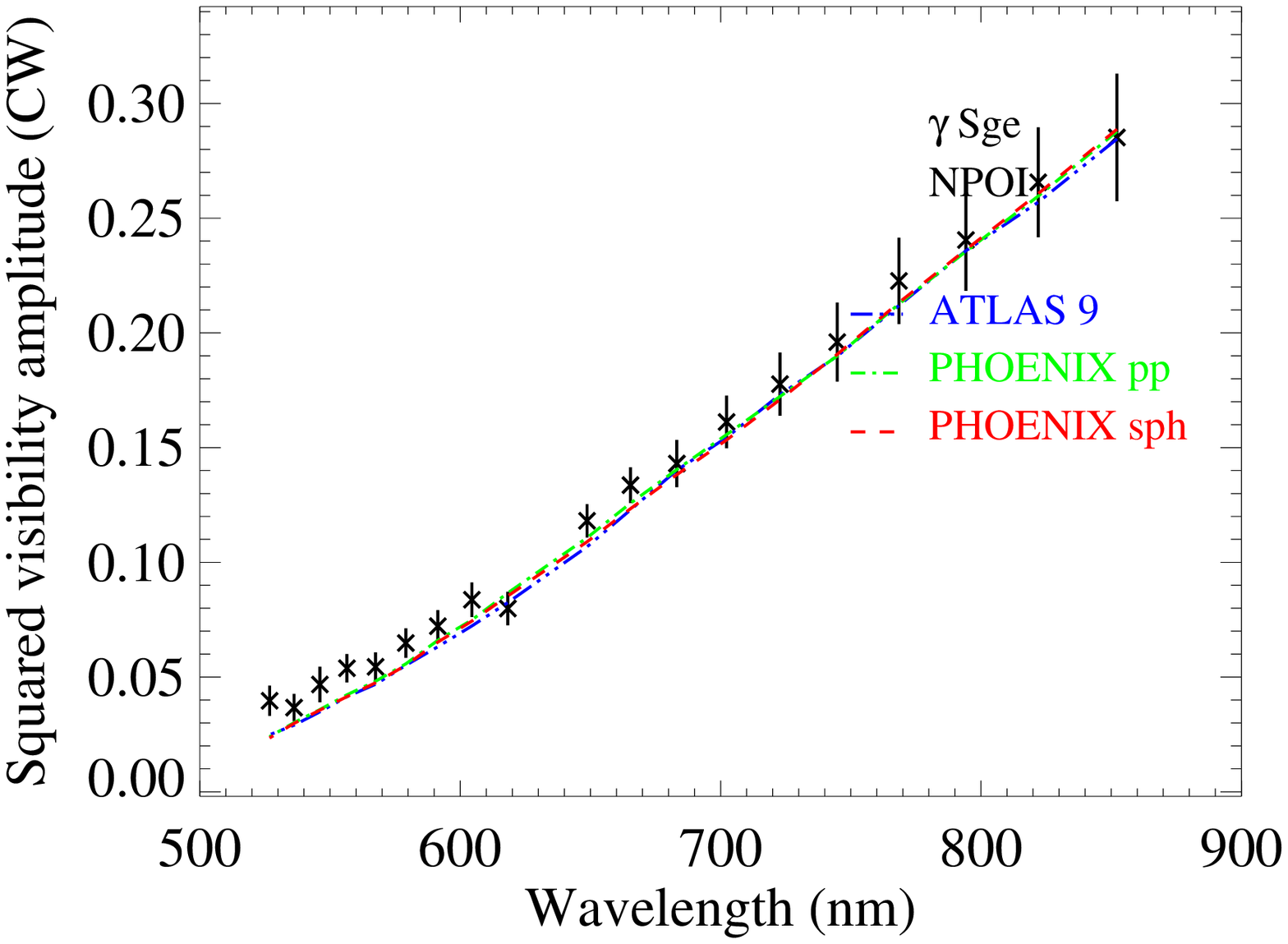}}
\resizebox{0.32\hsize}{!}{\includegraphics{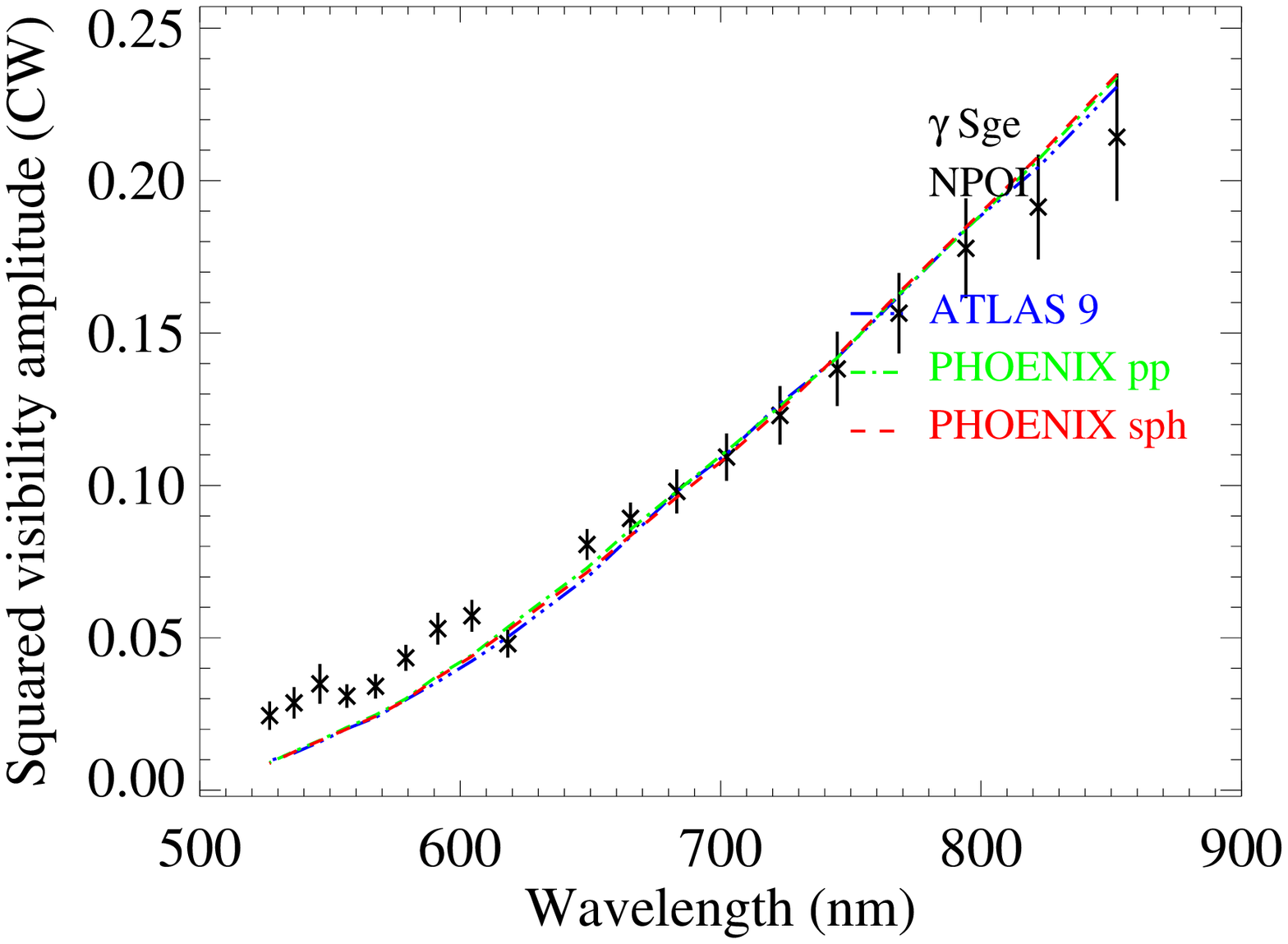}}

\resizebox{0.32\hsize}{!}{\includegraphics{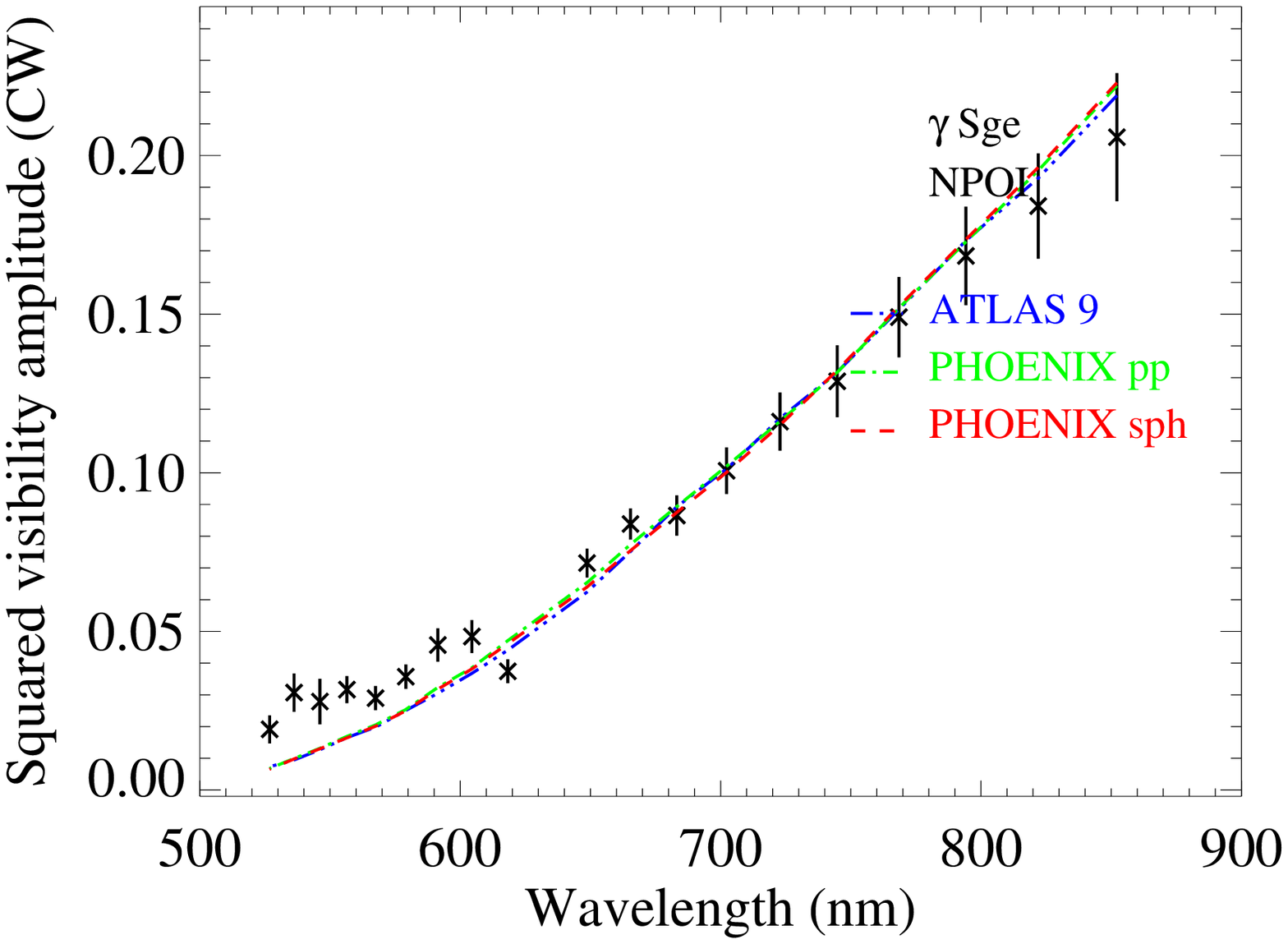}}
\resizebox{0.32\hsize}{!}{\includegraphics{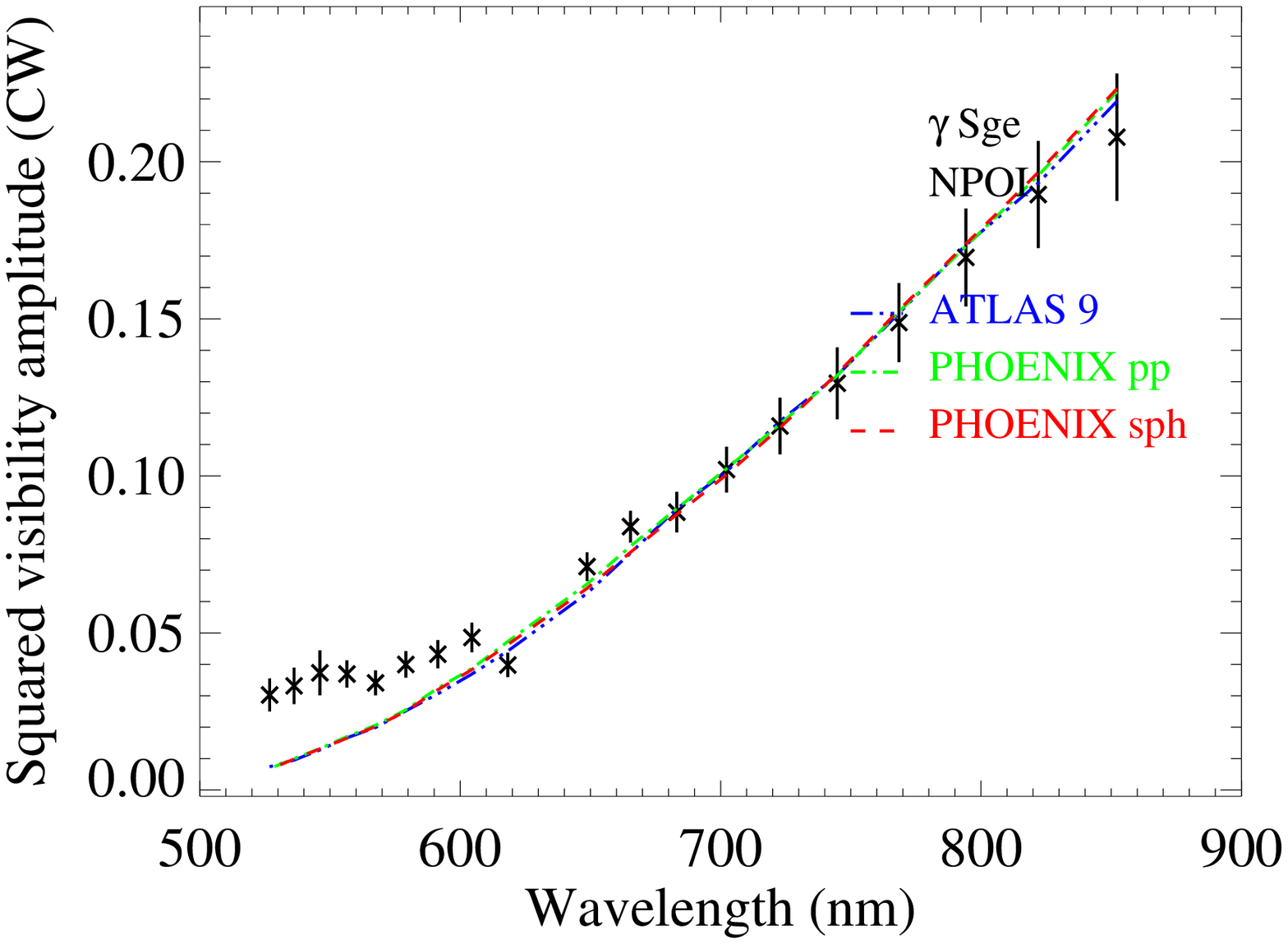}}
\resizebox{0.32\hsize}{!}{\includegraphics{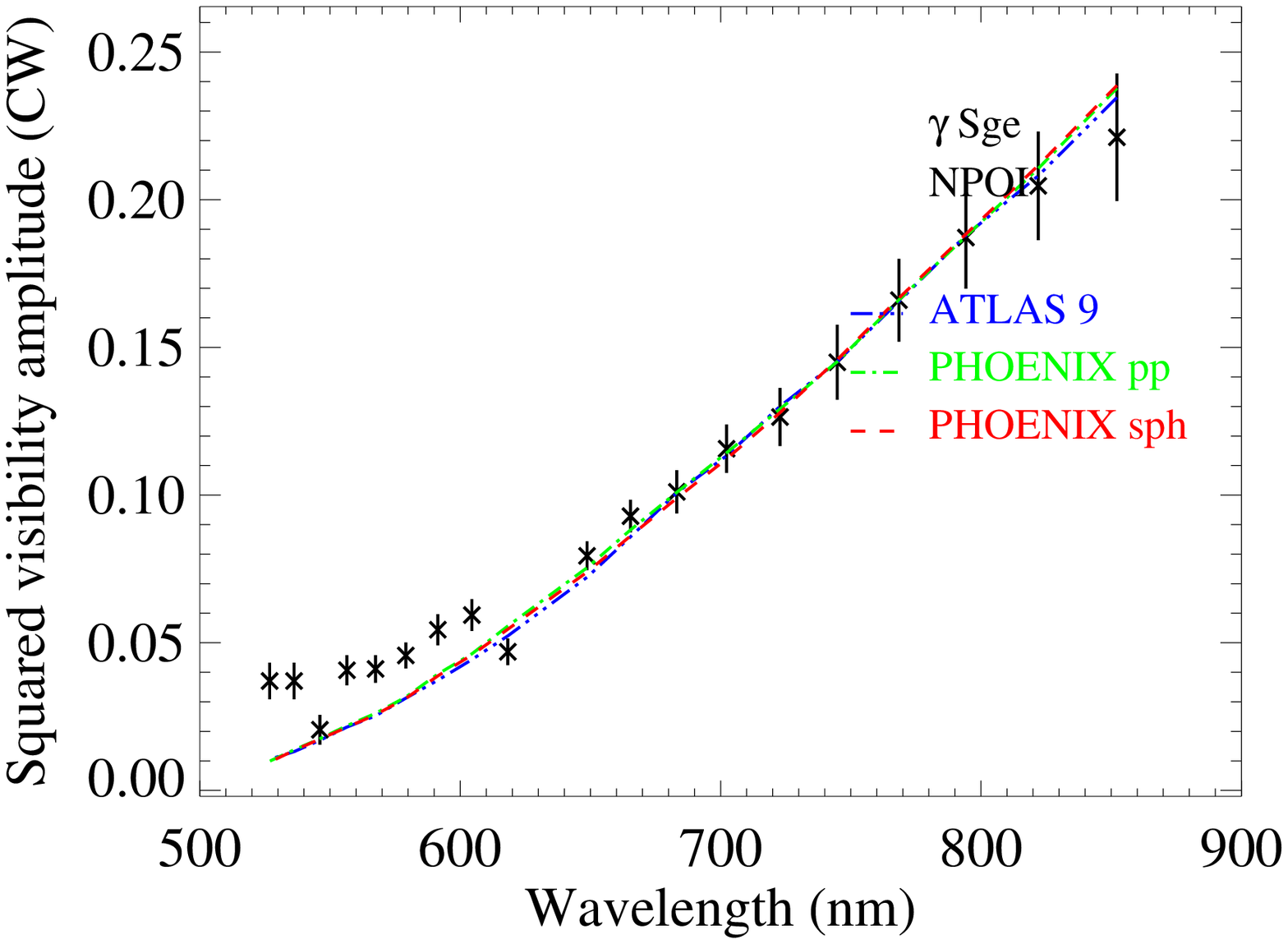}}
\caption{As Fig. \protect\ref{fig:npoivis1}, but showing the 
squared visibility amplitudes on the NPOI CW baseline.}
\label{fig:npoivis2}
\end{figure*}
\begin{figure*}
\centering
\resizebox{0.32\hsize}{!}{\includegraphics{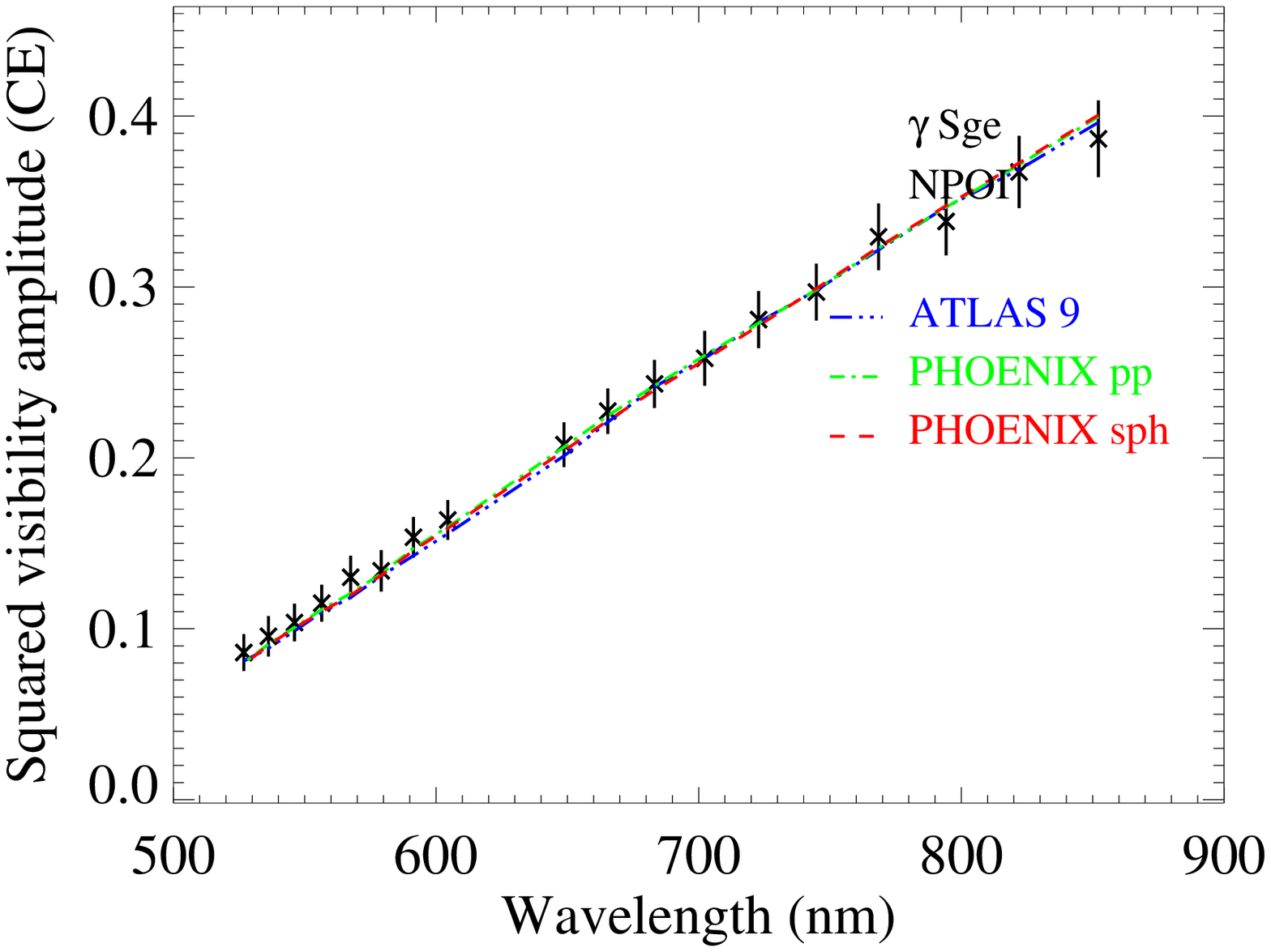}}
\resizebox{0.32\hsize}{!}{\includegraphics{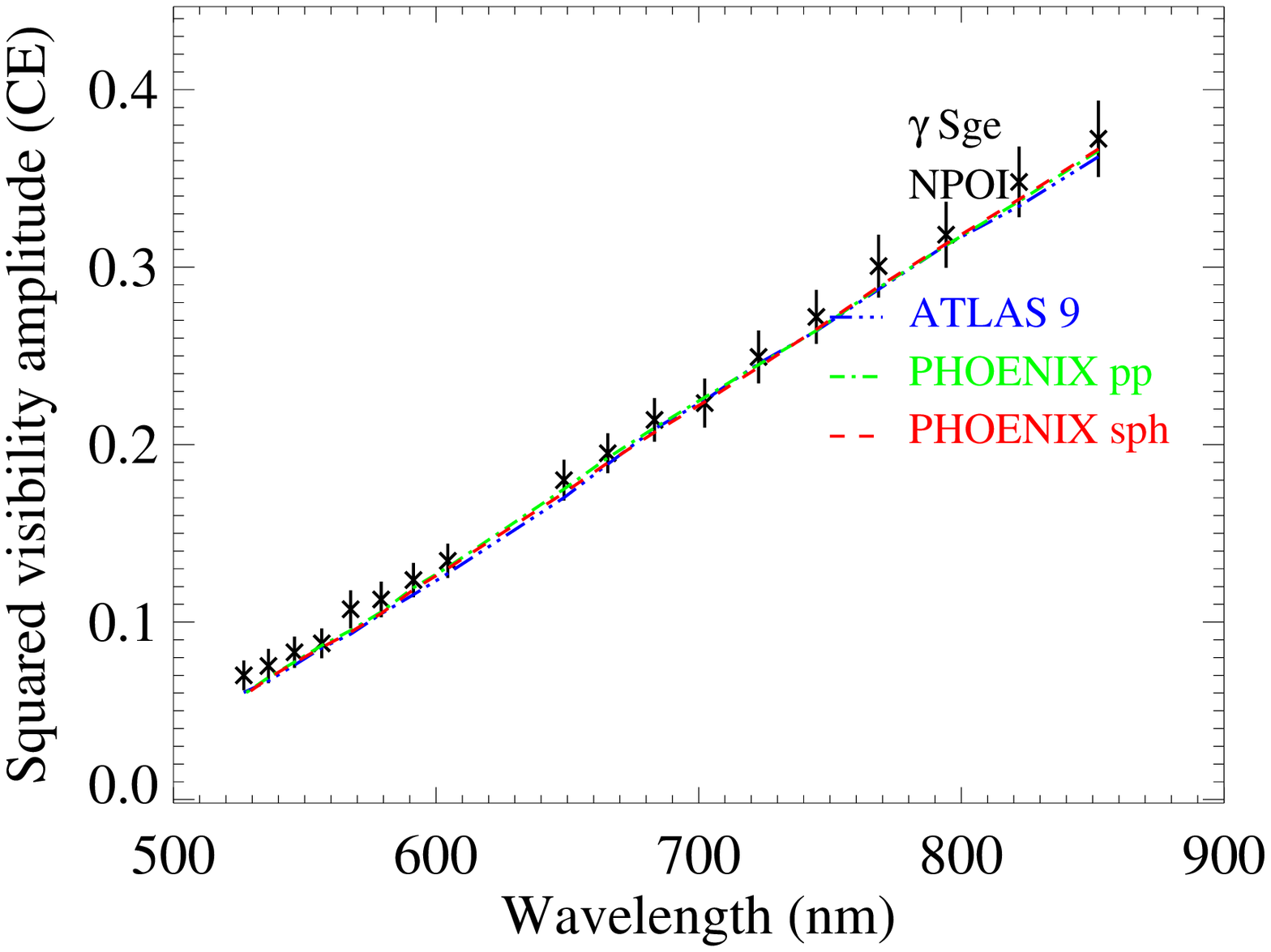}}
\resizebox{0.32\hsize}{!}{\includegraphics{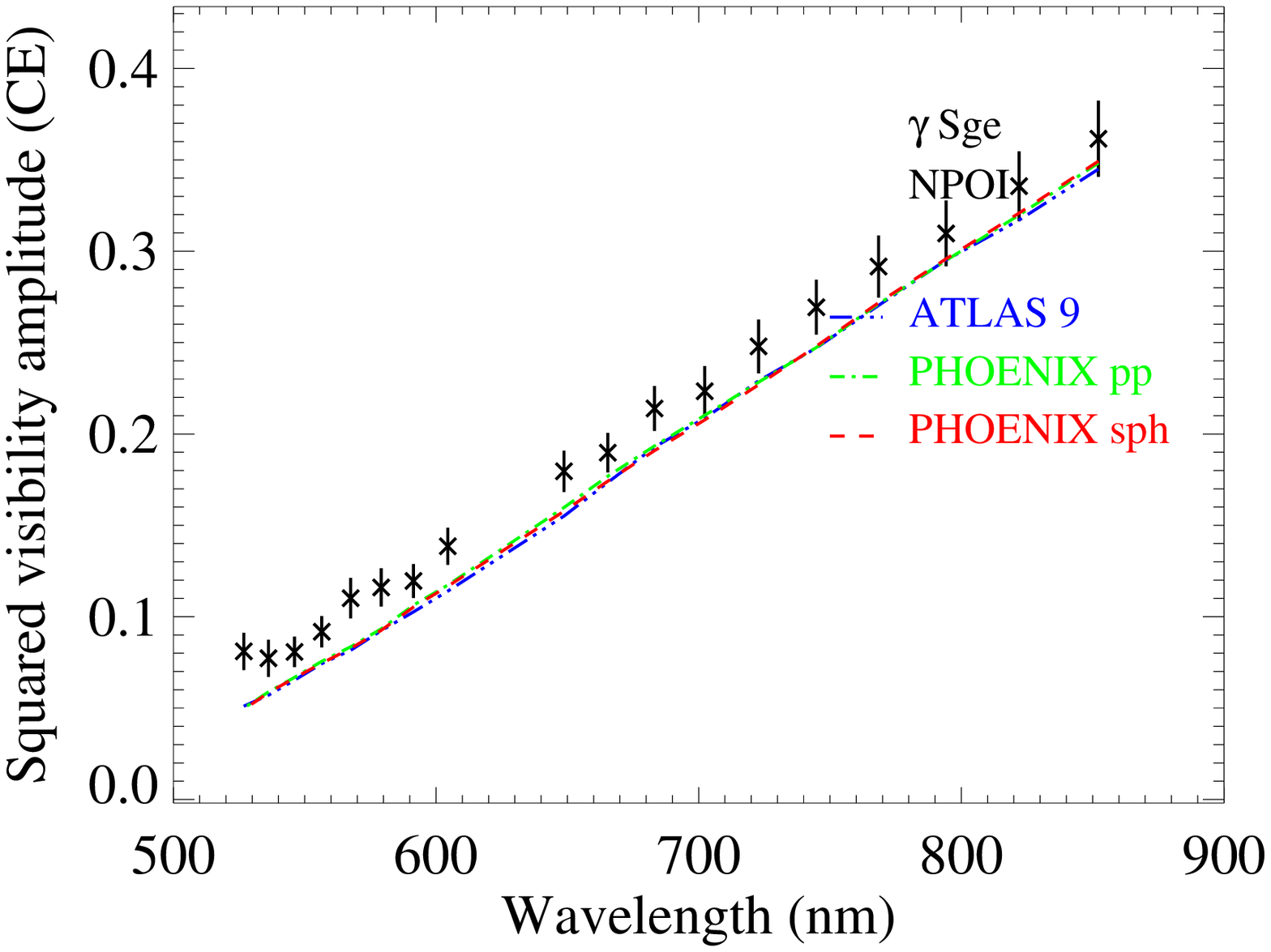}}

\resizebox{0.32\hsize}{!}{\includegraphics{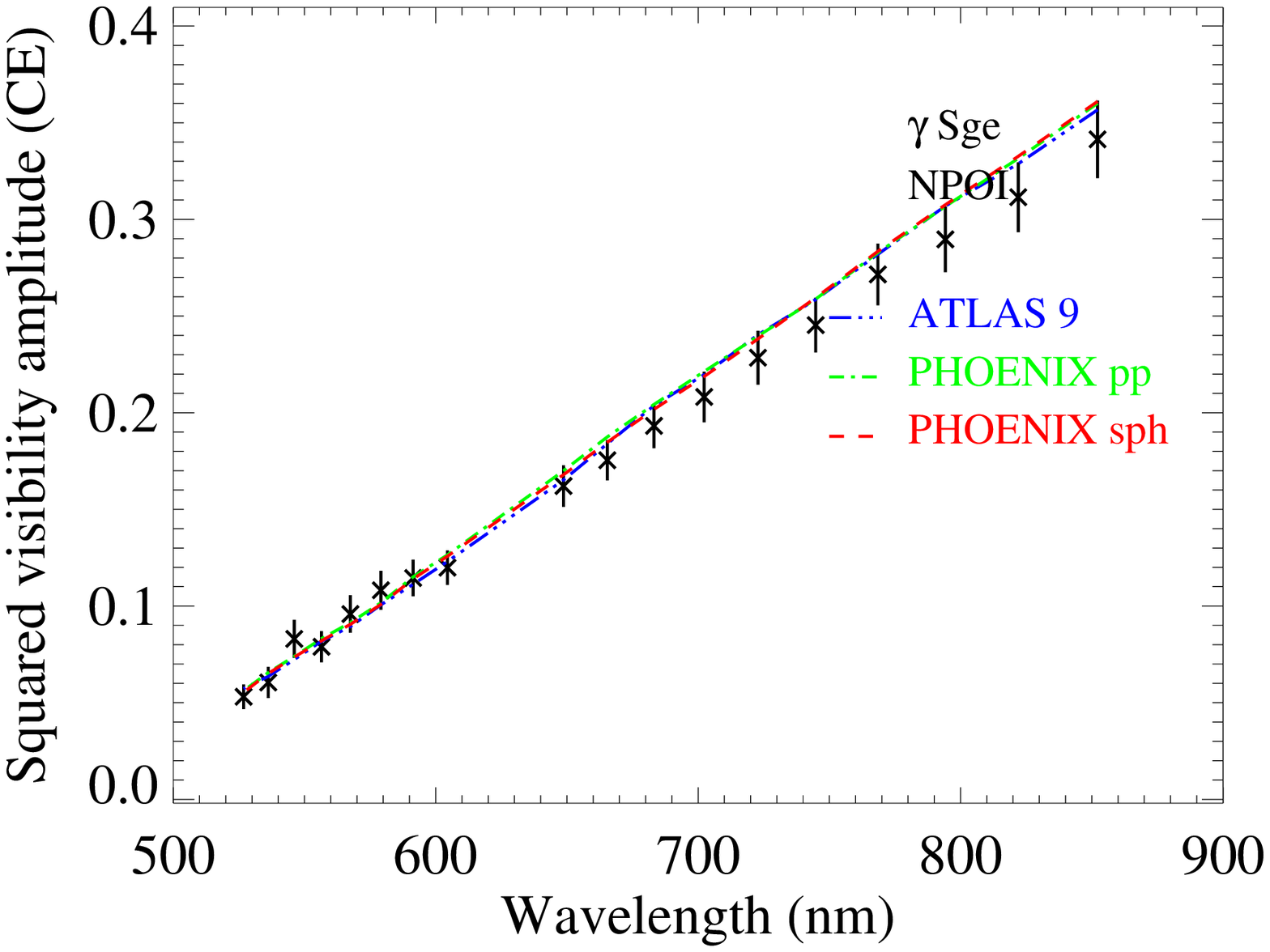}}
\resizebox{0.32\hsize}{!}{\includegraphics{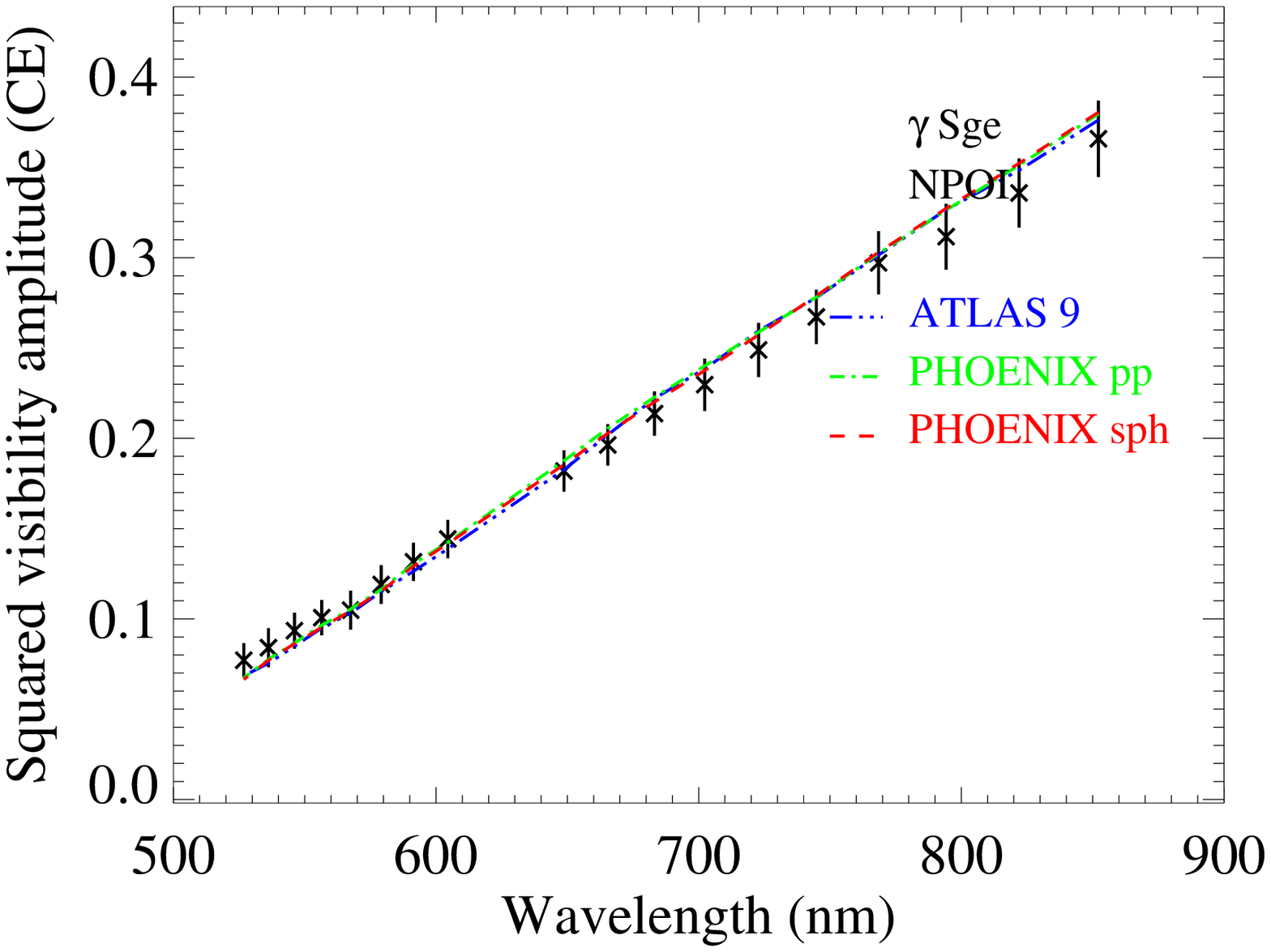}}
\resizebox{0.32\hsize}{!}{\includegraphics{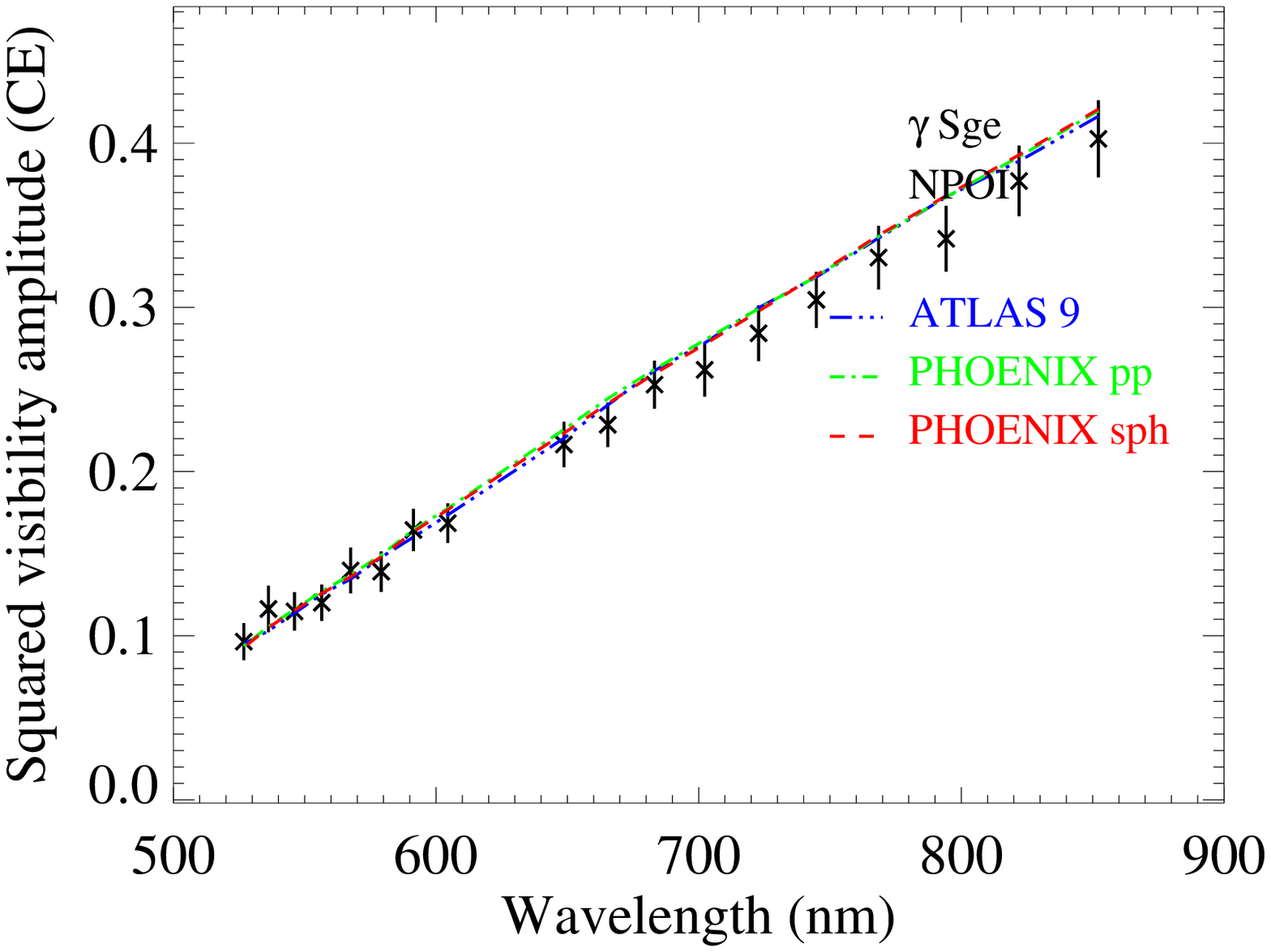}}
\caption{As Fig. \protect\ref{fig:npoivis1}, but showing the squared 
visibility amplitudes on the NPOI CE baseline.}
\label{fig:npoivis3}
\end{figure*}
\begin{figure*}
\centering
\resizebox{0.32\hsize}{!}{\includegraphics{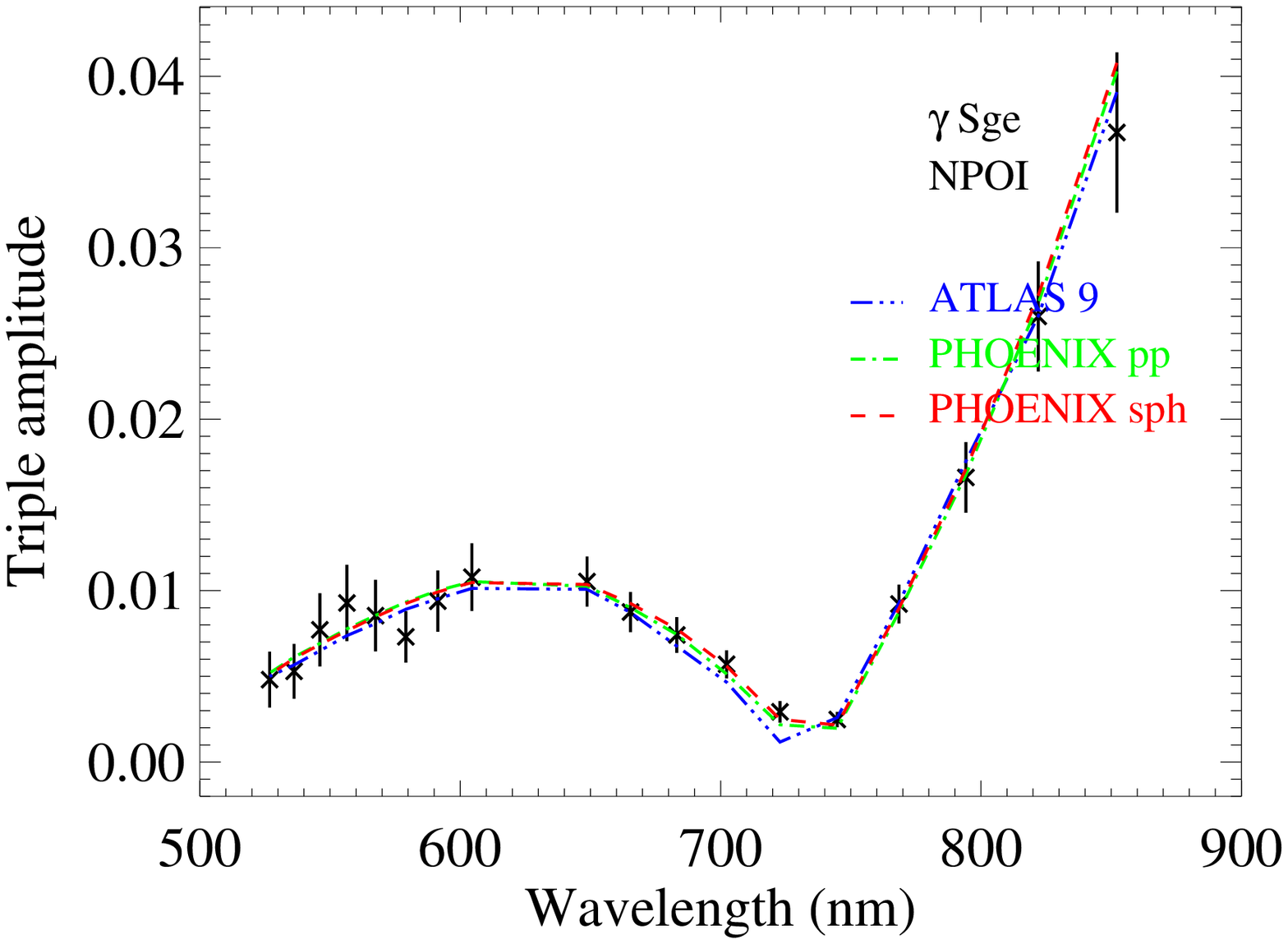}}
\resizebox{0.32\hsize}{!}{\includegraphics{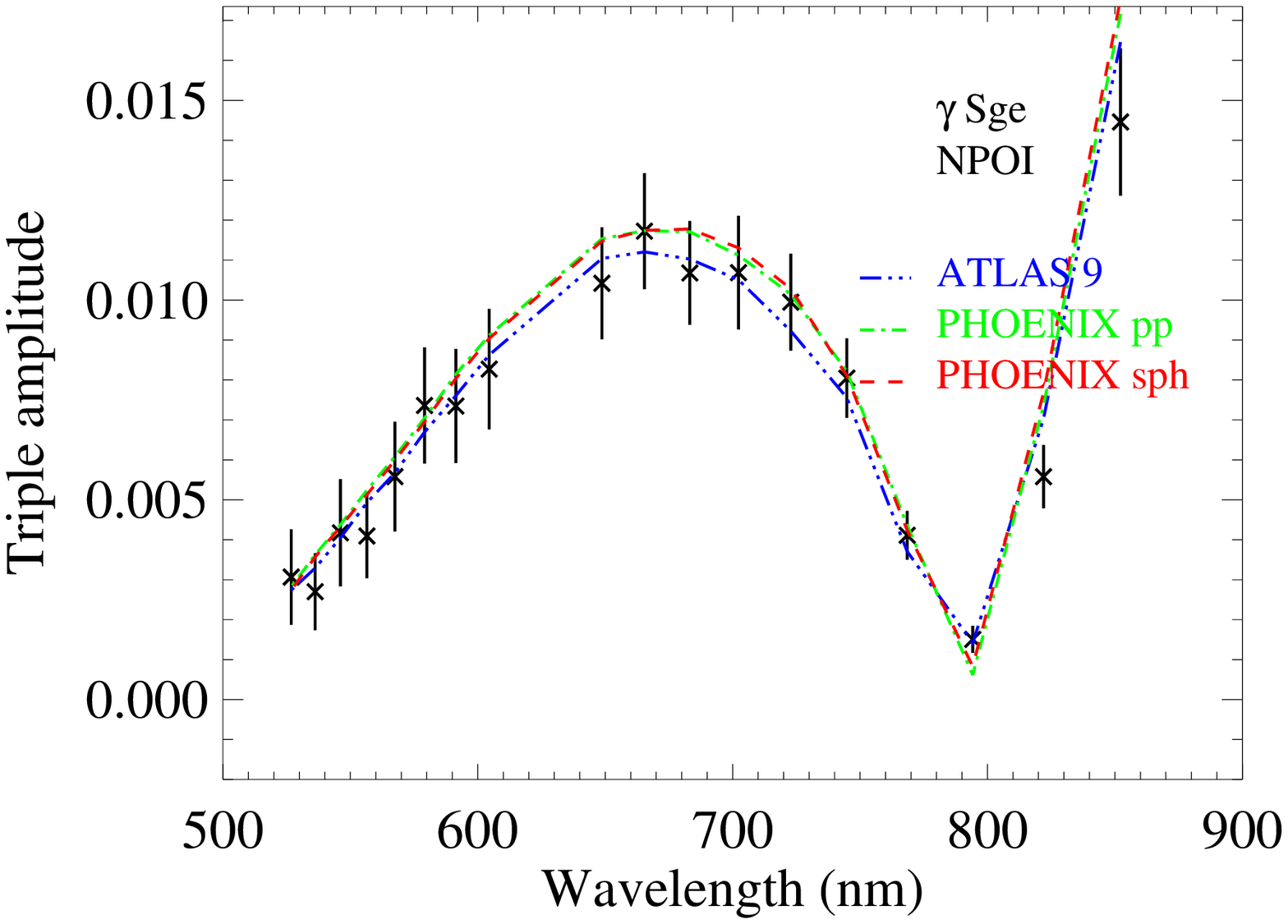}}
\resizebox{0.32\hsize}{!}{\includegraphics{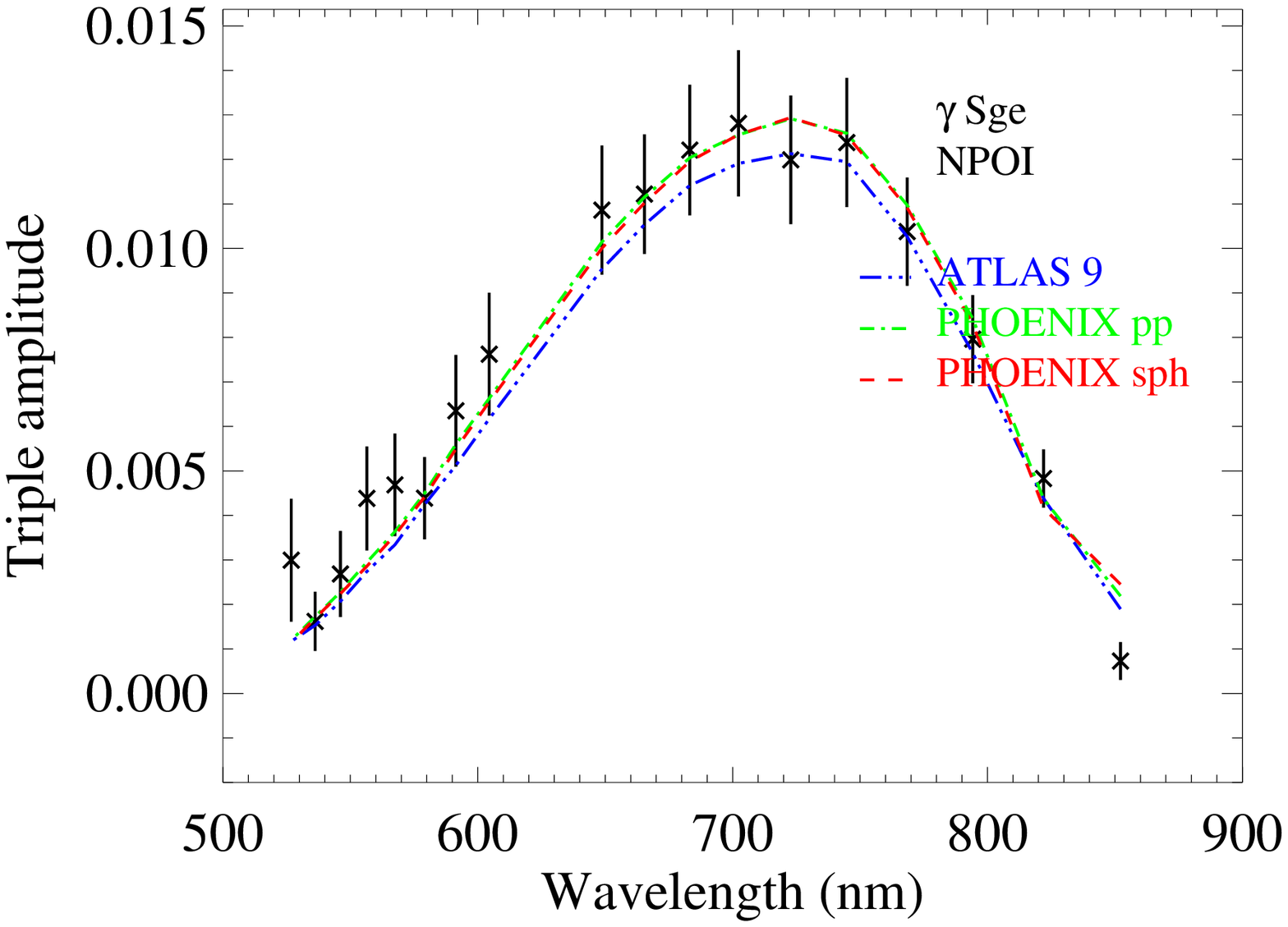}}

\resizebox{0.32\hsize}{!}{\includegraphics{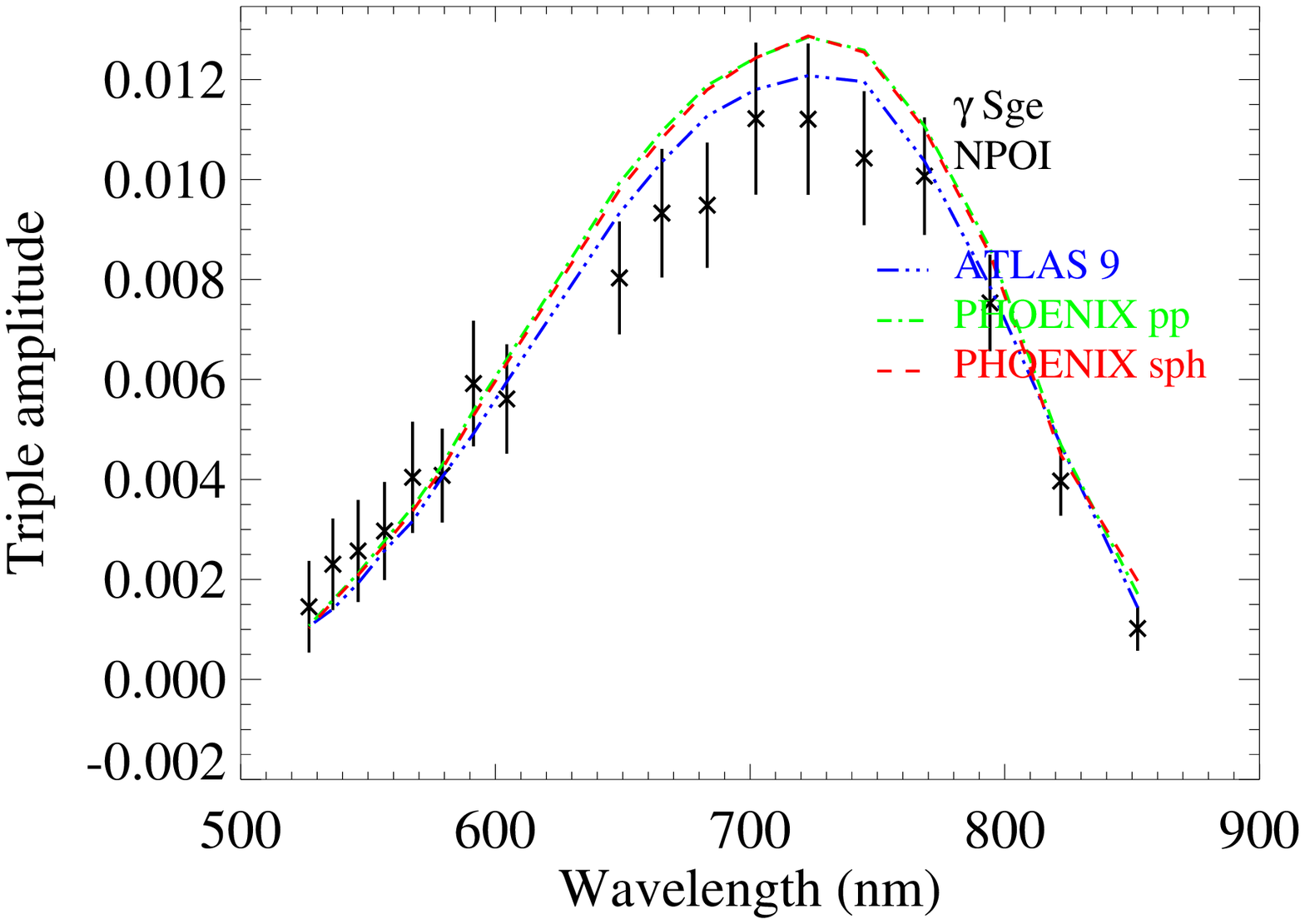}}
\resizebox{0.32\hsize}{!}{\includegraphics{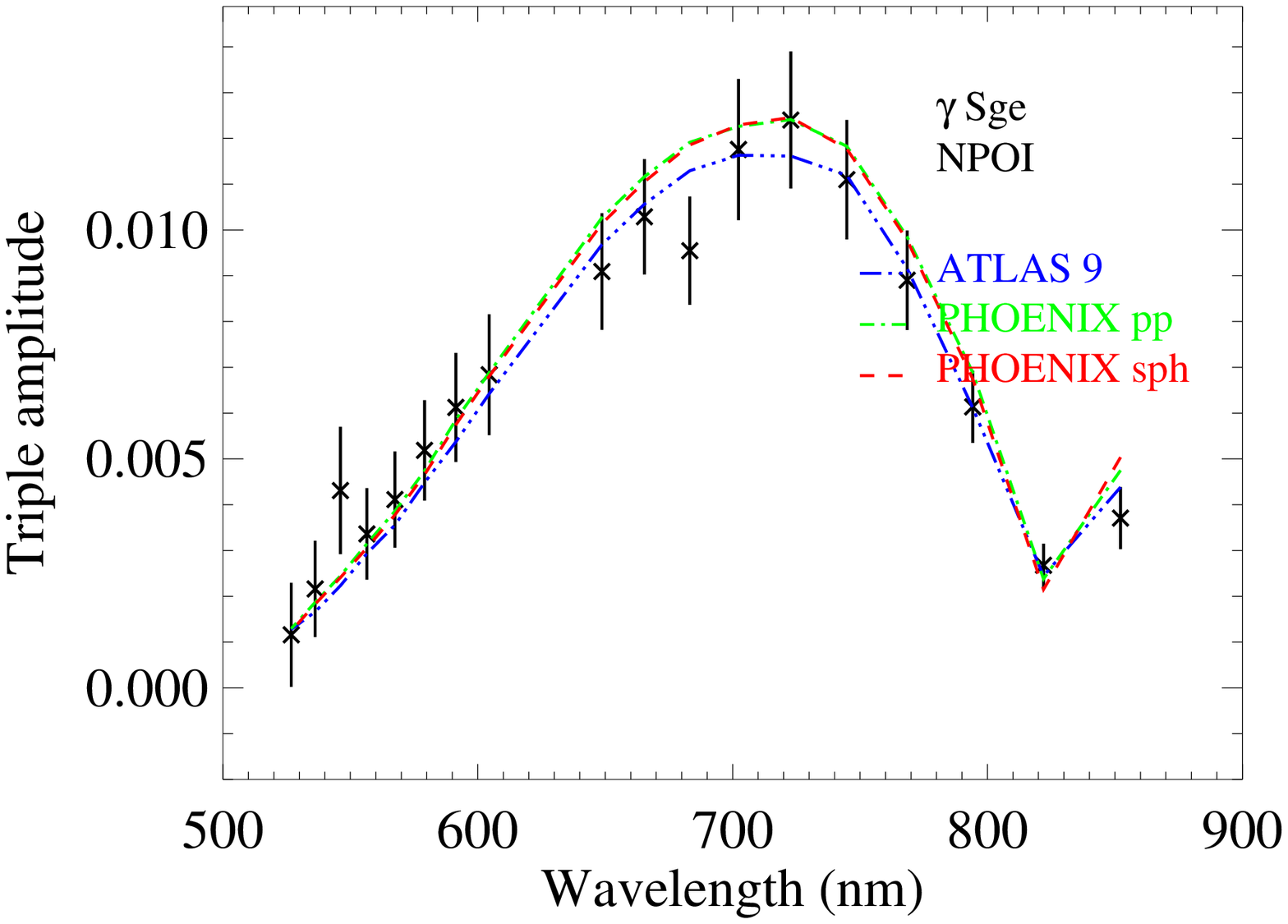}}
\resizebox{0.32\hsize}{!}{\includegraphics{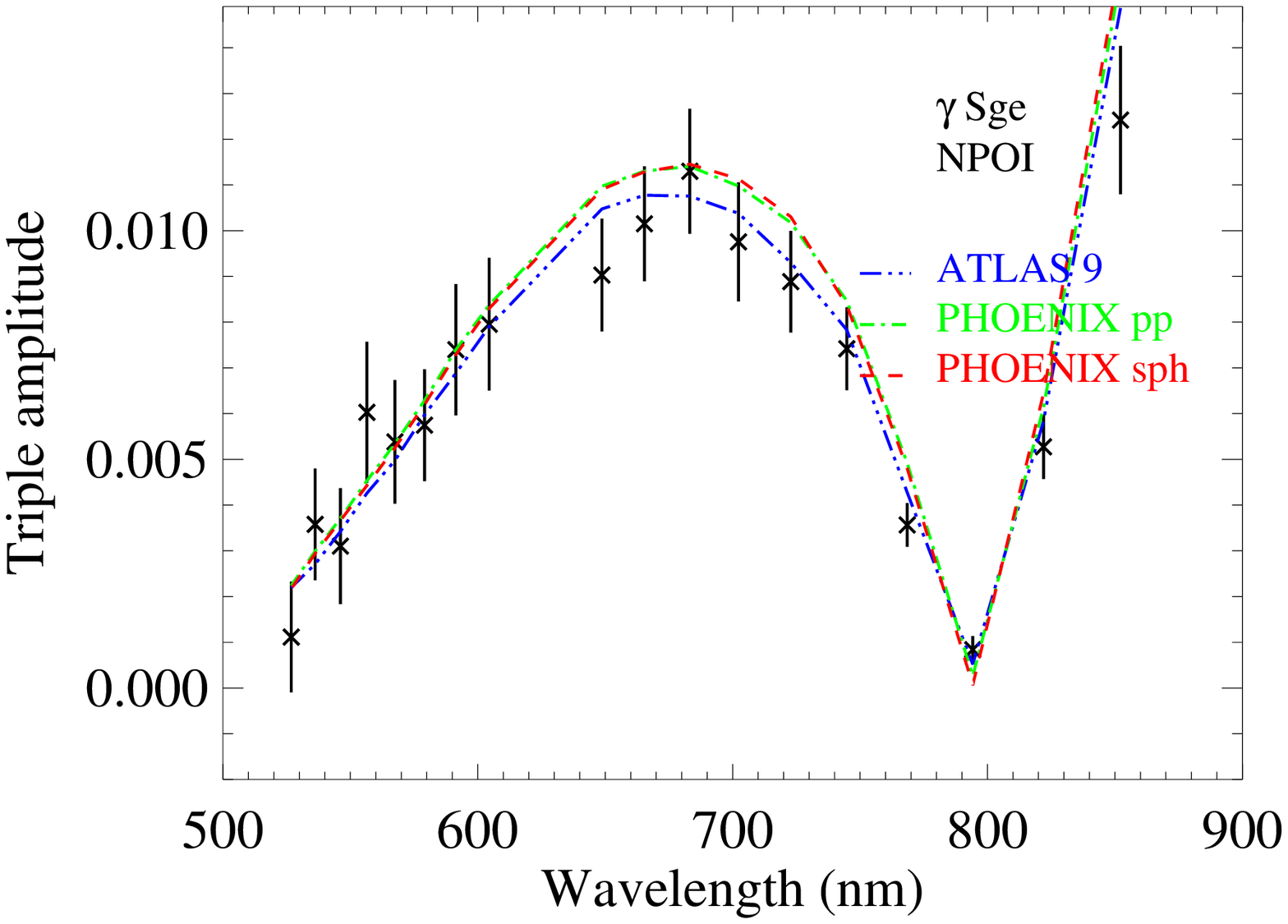}}
\caption{As Fig. \protect\ref{fig:npoivis1}, but showing the
NPOI triple amplitudes.}
\label{fig:npoiampl}
\end{figure*}
\begin{figure*}
\centering
\resizebox{0.32\hsize}{!}{\includegraphics{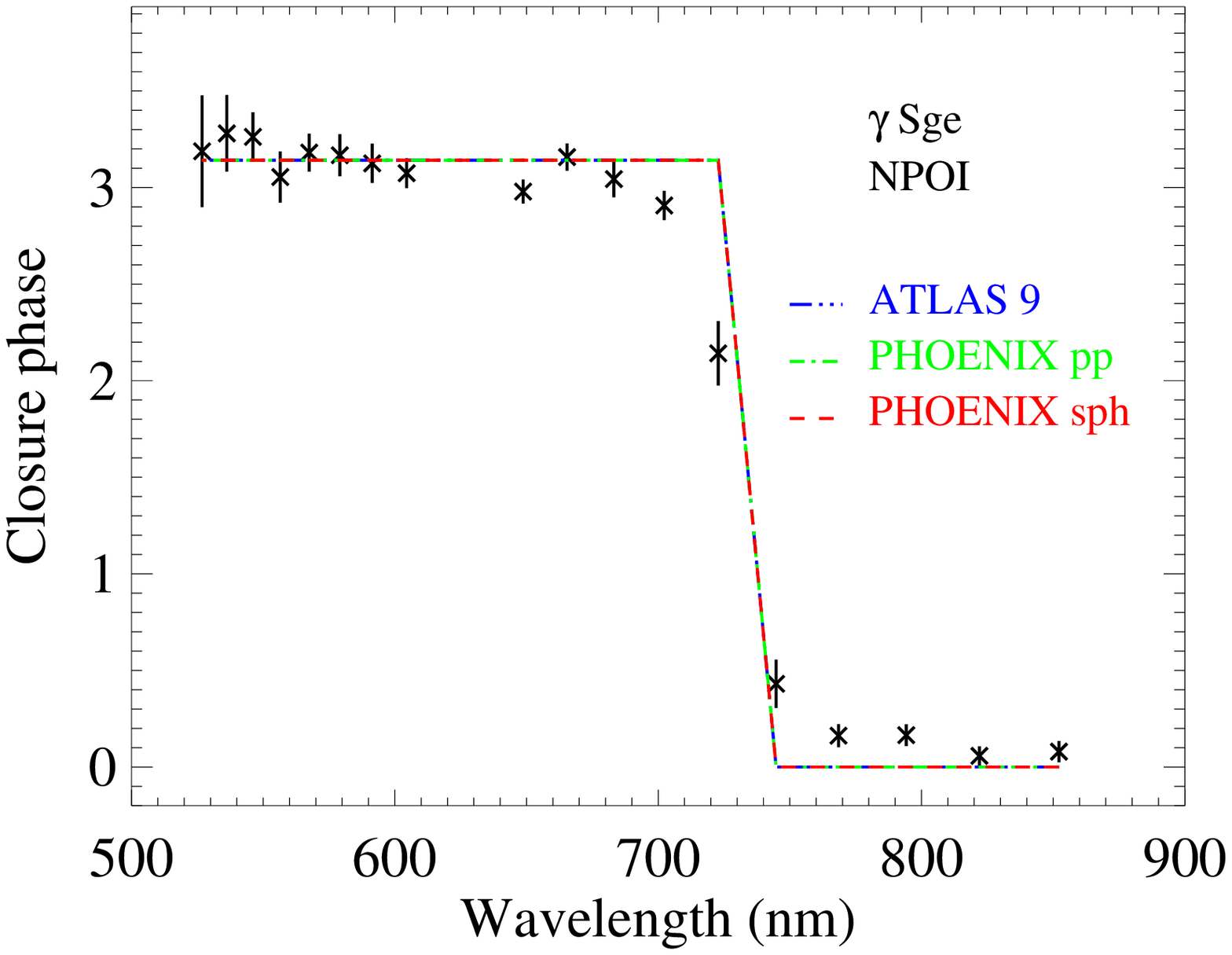}}
\resizebox{0.32\hsize}{!}{\includegraphics{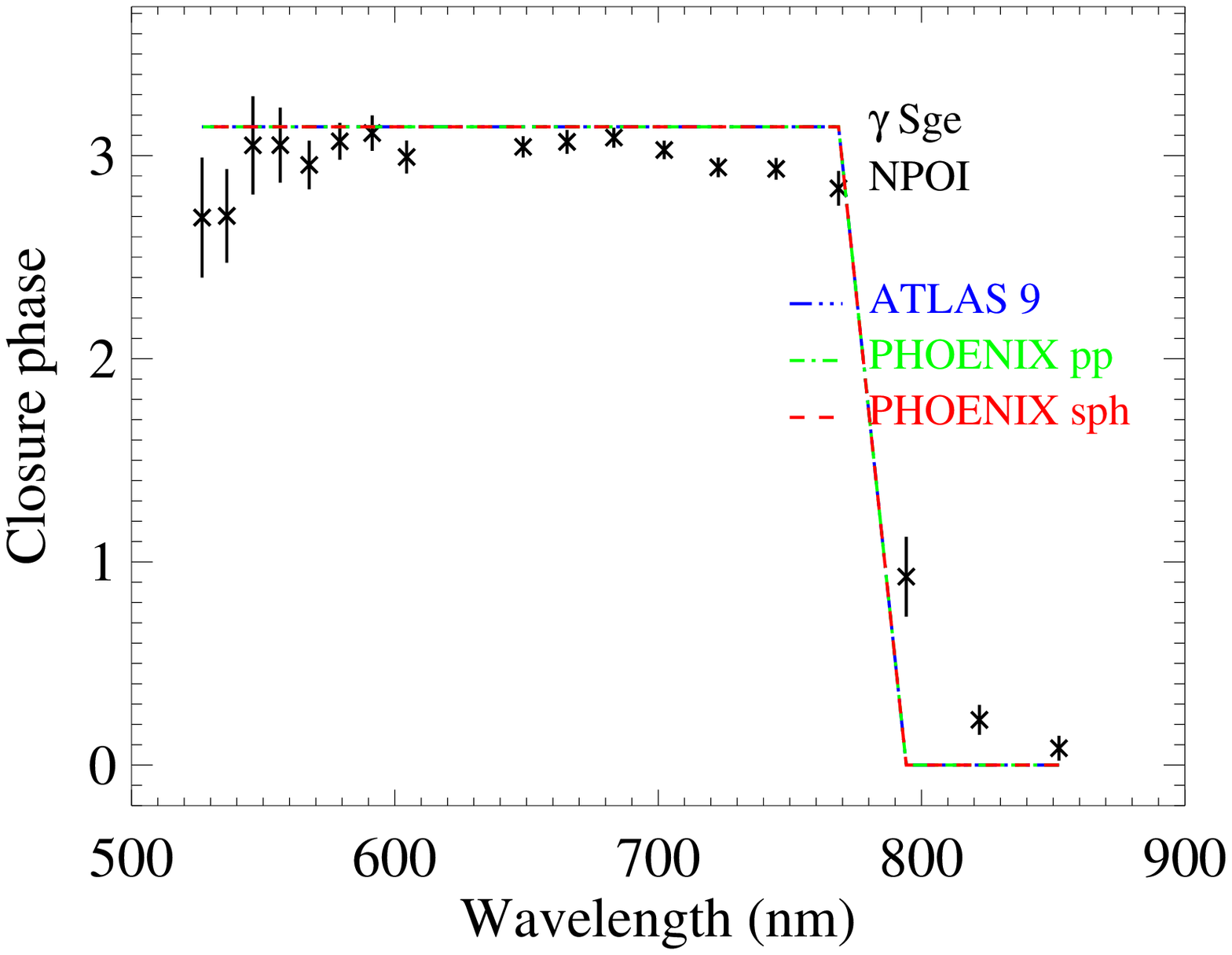}}
\resizebox{0.32\hsize}{!}{\includegraphics{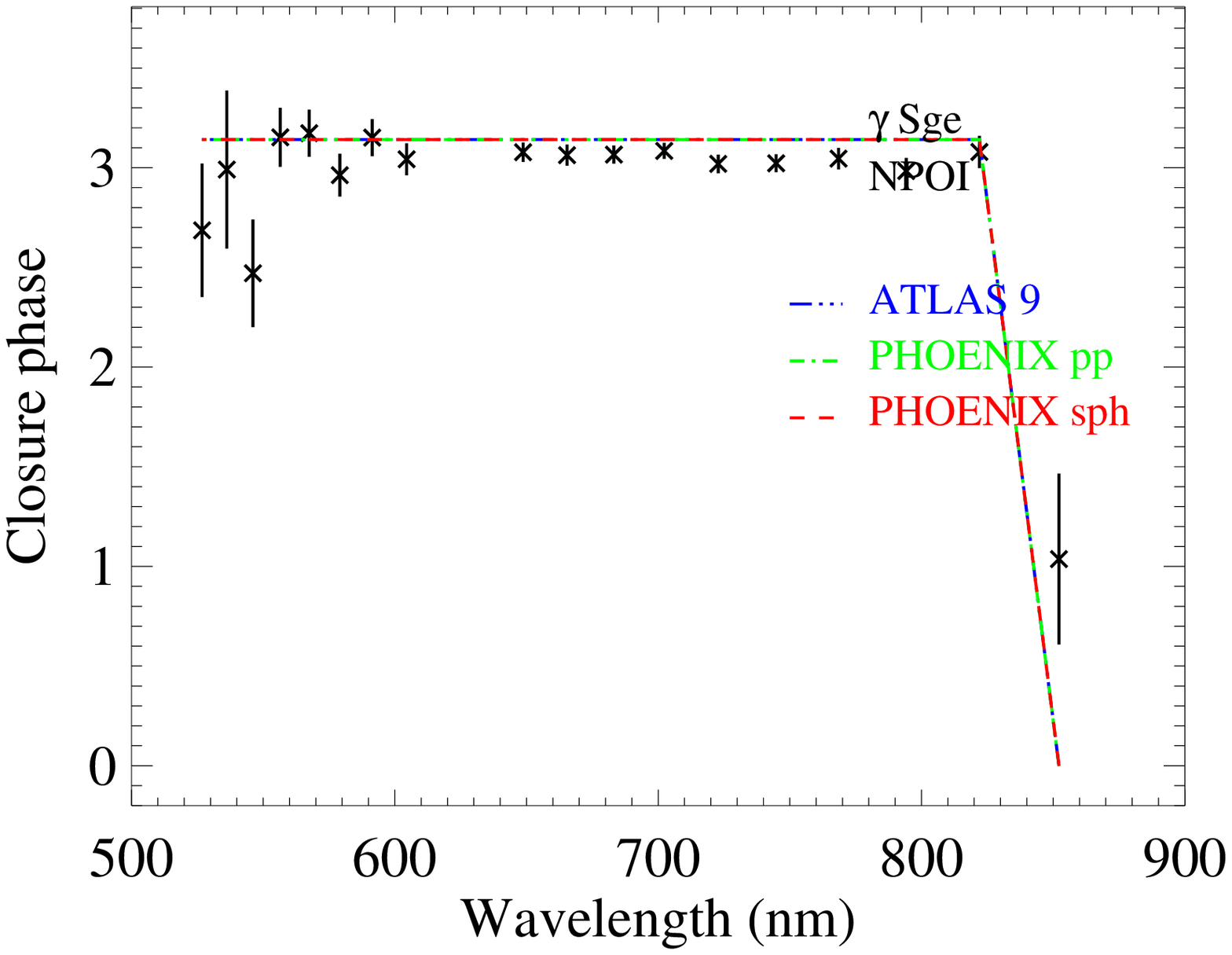}}

\resizebox{0.32\hsize}{!}{\includegraphics{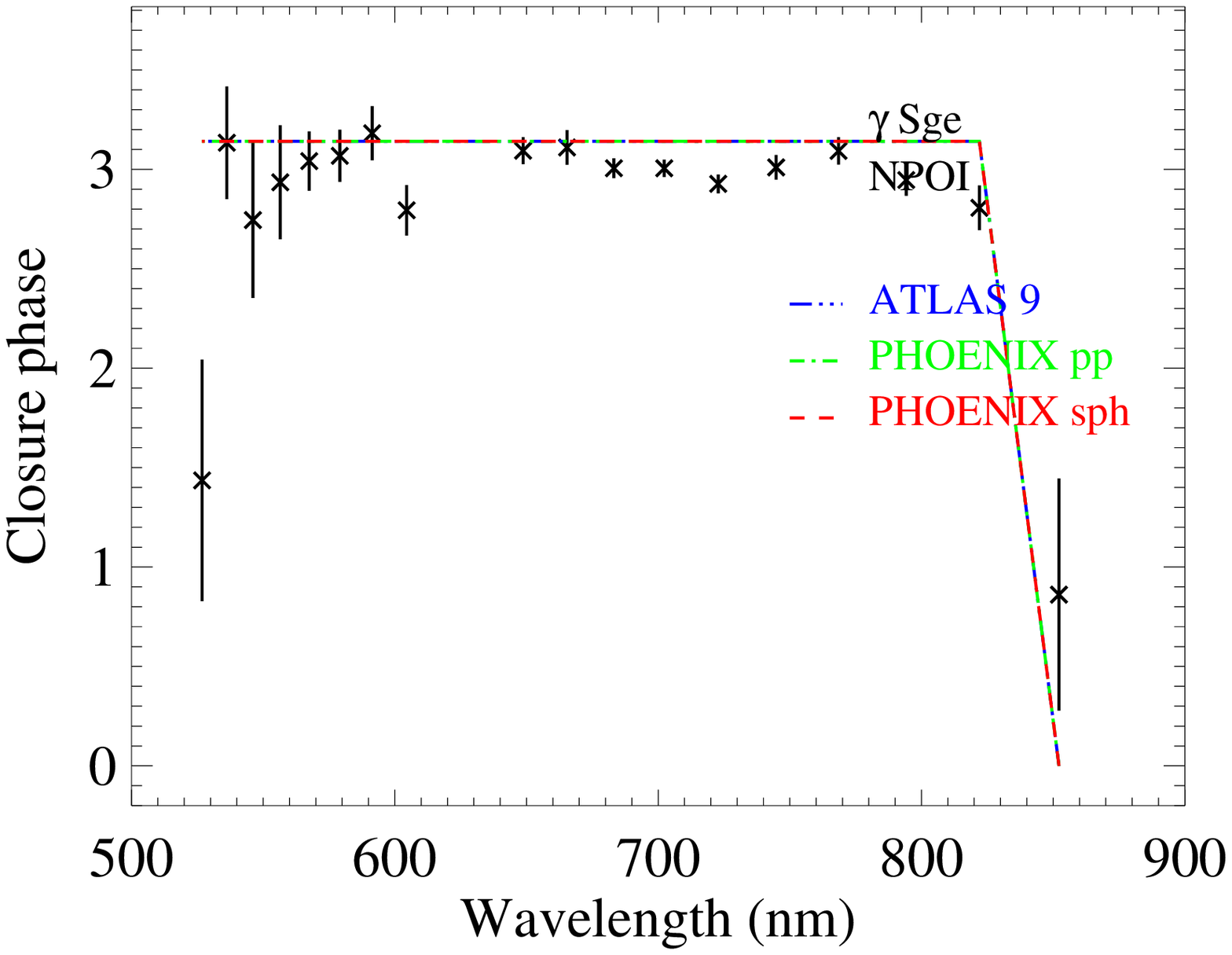}}
\resizebox{0.32\hsize}{!}{\includegraphics{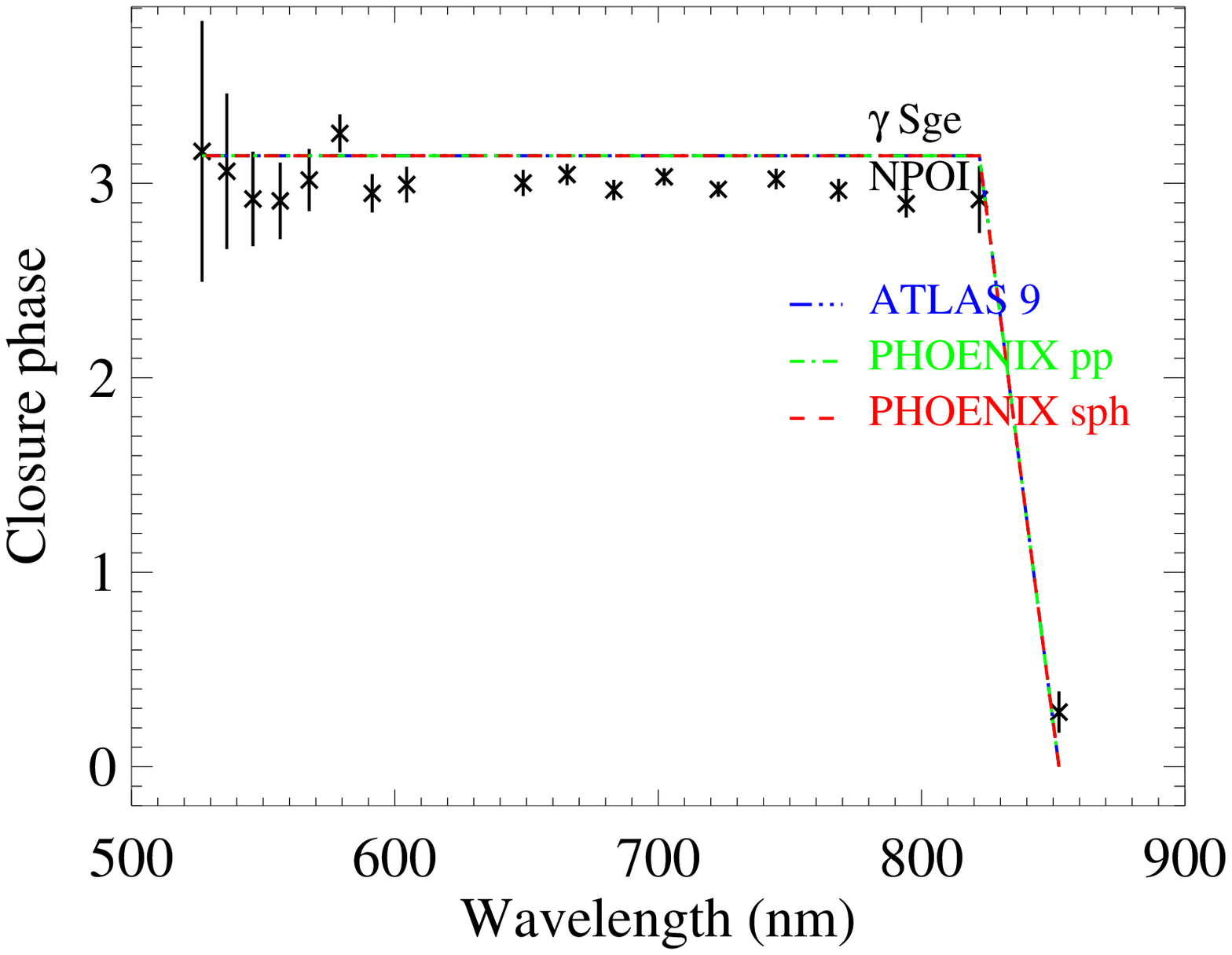}}
\resizebox{0.32\hsize}{!}{\includegraphics{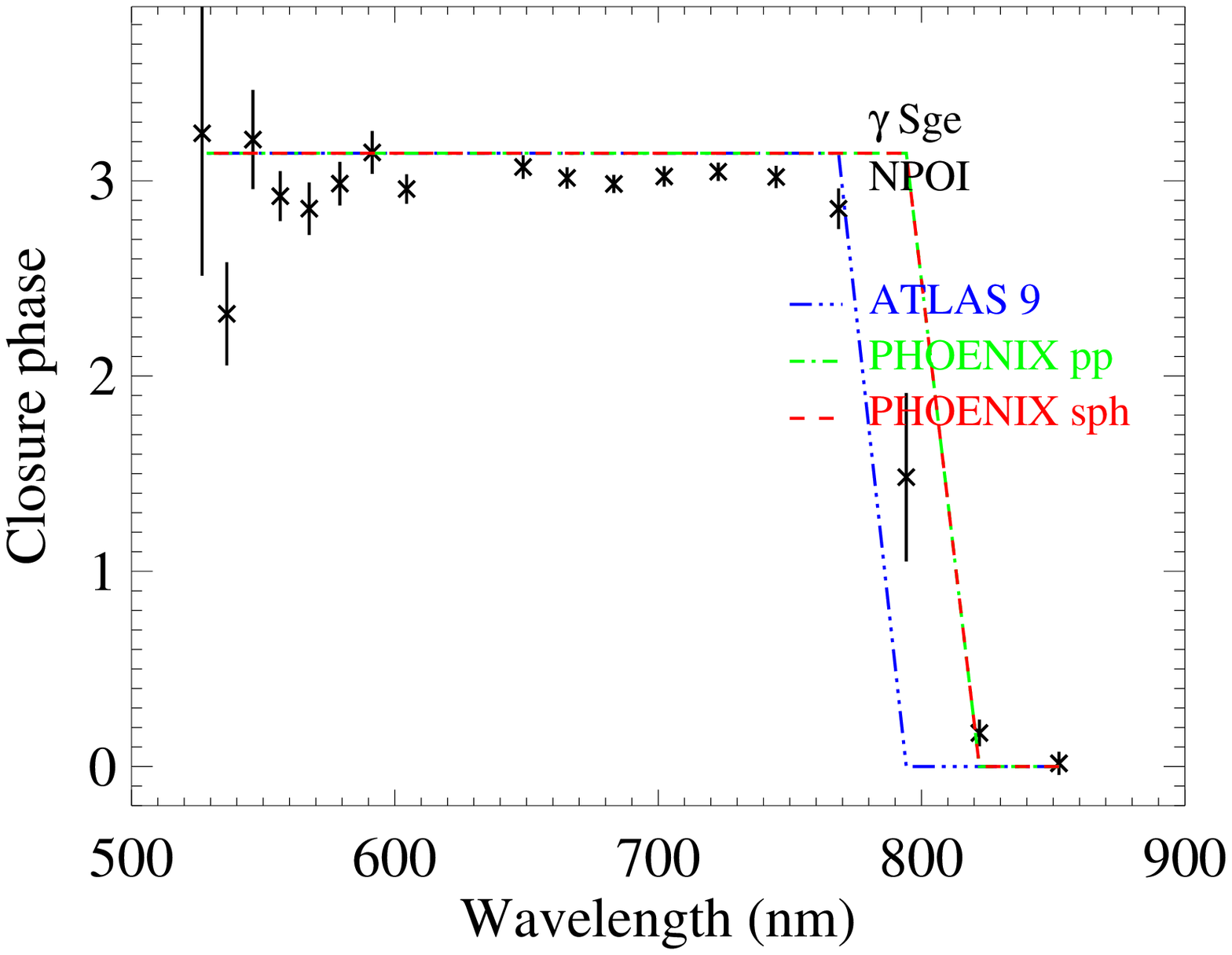}}
\caption{As Fig. \protect\ref{fig:npoivis1}, but showing the
NPOI closure phases. Note that the slope of the model flip from 0 to $\pi$
is an artifact because the model values are only calculated for 
each NPOI spectral channel.}
\label{fig:npoiphas}
\end{figure*}
\subsection{NPOI results}
\label{sec:npoiresults}
As a first characterisation of our NPOI data, we use models of 
a uniform disc (UD, $I=1$ for  $0\le\mu\le1$, $I=0$ otherwise),
and a fully darkened disc (FDD, $I=\mu$).
Here, $I$ is the intensity, $\mu=\cos\Theta$ (or $\mu=\sqrt{1-(r/R)^2}$) 
the cosine of the angle between the line of sight and the normal
of the surface element of the star ($R$ the stellar radius, $r$ the 
distance from the centre of the disc). 
Monochromatic synthetic visibility values $V$ were obtained 
for the UD and FDD cases, and subsequently integrated over the bandpass
of each NPOI spectral channel (covering frequencies $\nu_1$ to $\nu_2$)
as
\begin{equation}
\label{eq:broadnpoi}
V_i=\frac{\int_{\nu_1}^{\nu_2}\,F_\nu\,V(\nu)\,d\nu}{\int_{\nu_1}^{\nu_2} F_\nu\,d\nu}
\end{equation}
with $F_\nu$ the flux from the blackbody radiation. We use 
$T_\mathrm{eff}=3768$\,K
(see Sect.~\ref{sec:introduction}). Variations of $T_\mathrm{eff}$ within its
uncertainties do not have a significant effect on the visibility values.
Finally, the synthetic squared visibility values for each of the baselines 
($|V_i(CE)|^2$, $|V_i(CW)|^2$, $|V_i(EW)|^2$), the triple amplitude
($V_i^{CEW}=|V_i(CE)\, V_i(CW)\, V_i(EW)|$), and the closure phase 
($\Phi_i^{CEW}=0$ if $V_i(CE)\, V_i(CW)\, V_i(EW) > 0$,  
$\Phi_i^{CEW}=\pi$ if $V_i(CE)\, V_i(CW)\, V_i(EW) < 0$) were obtained.

Note that in the case of NPOI the monochromatic visibility amplitudes are
integrated before building the square, while for VLTI/VINCI the 
monochromatic {\it squared} visibility 
amplitudes are integrated (cf. Paper\,II). The reason for the difference
is that the data processing of NPOI first integrates the photons on the APDs 
and the squared visibility is computed from the already integrated bin counts, 
while for  VLTI/VINCI first the full powerspectrum is computed and
integrated thereafter. 
This leads to noticeable differences, in particular around the minima 
of the visibility function.

Best fitting angular diameters $\Theta_\mathrm{UD,FDD}$ 
are derived from a least square optimisation. The resulting
values are shown in Table~\ref{tab:alphas} together with the 
reduced $\chi^2_\nu$ values. The number of degrees of freedom is
644 (7 observations times 19 spectral channels times (3 squared visibility
plus 1 triple amplitude plus 1 closure phase), minus 21 values flagged
for quality).
\begin{table}
\centering
\caption{Fit results of our NPOI data to models of a uniform disc (UD)
and a fully darkened (FDD) disc.
The formal errors of the diameter values
are $\sim 0.01$\,mas, additional calibration uncertainties are
$\sim 0.06$\,mas, total errors thus $\sim 0.06$\,mas.}
\label{tab:alphas}
\begin{tabular}{lrrr}
\hline\hline
Model & Diameter & Parameter $\alpha$ & $\chi^2_\nu$ \\\hline
UD             & $\Theta_\mathrm{UD}=5.64$\,mas  & $\alpha=0$ & 11.0\\
FDD            & $\Theta_\mathrm{FDD}=6.59$\,mas & $\alpha=1$ & 5.6 \\\hline
\end{tabular}
\end{table}
The formal errors of the obtained diameter values are of the order of
0.01\,mas, and are small compared to 
calibration uncertainties that are estimated to $\sim$\,1\%$\sim$0.06\,mas.
Total errors are thus $\sim 0.06$\,mas.

Figures~\ref{fig:npoivis1}, \ref{fig:npoivis2}, \ref{fig:npoivis3}, 
\ref{fig:npoiampl}, and \ref{fig:npoiphas} show the obtained NPOI squared 
visibility amplitudes on baselines EW, CE, CW, the NPOI triple amplitudes, 
and the NPOI closure phases, respectively. Also shown 
are the model atmosphere predictions
as described below in Sect.~\ref{sec:models}.

The gain with respect to Paper\,I in the signal-to-noise ratio and in 
the number of usable spectral channels is thanks to the method of 
coherent integration, as can be seen by comparing these Figs.
to the results based on incoherent averaging (Fig.~3 of Paper\,I).

The results (Figs.~\ref{fig:npoivis1}-\ref{fig:npoiphas} and Table~\ref{tab:alphas}) show that the visual intensity profile of $\gamma$\,Sge
is limb-darkened, clearly closer to a FDD model than to a UD model, while
both of these simple descriptions do not provide a very good representation 
of our data.
A detailed comparison of our visibility data to model atmosphere predictions 
is discussed below in Sect.~\ref{sec:models}.
\section{VLTI/VINCI measurements}
\subsection{VLTI/VINCI observations}
The near-infrared $K$-band interferometric data of $\gamma$ Sagittae were
obtained with the ESO Very Large Telescope Interferometer 
(VLTI, Glindemann et al. \cite{glindemann03}), the
instrument VINCI (Kervella et al. \cite{kervella03}), and the two
VLTI test siderostats on June 28, July 8, July 11, July 15, August 8,
September 12, and September 18, 2002. These data are public commissioning
data released from the 
VLTI\footnote{http://www.eso.org/projects/vlti/instru/vinci/vinci\_data\_sets.html}. 
The VLTI stations E0 and G1 forming a ground baseline length of 66\,m were 
used for all our observations.
The observations were repeated during 7 different nights spread over
more than 2 months in order to compute the night-to-night variation of 
the obtained diameter and thereby to estimate the calibration uncertainty 
caused by different atmospheric and possibly instrumental conditions.
All data were obtained as series of typically 100 or 500 interferograms
with a scan length of 250\,$\mu$m and a fringe frequency of 295\,Hz.

The stars \object{56\,Aquilae} and \object{31\,Orionis} were used as primary 
calibration stars and were observed in each of our observation nights 
close in time to the
$\gamma$\,Sagittae observations. A number of additional calibration stars
observed during these nights were used as secondary calibrators
for $\gamma$\,Sagittae. The characteristics of all calibration stars used
are taken from Bord\'e et al. (\cite{borde02}, 
based on Cohen et al. \cite{cohen99}) and are listed in 
Table~\ref{tab:calibrators}.
\begin{table}
\centering
\caption{Characteristics of the stars that were used as calibration stars
for our VLTI/VINCI observations of $\gamma$\,Sagittae. Listed are the spectral 
type, the $K$-band magnitude, the uniform-disc diameter and its error, and
the effective temperature, all from Bord\'e et al. (\cite{borde02}, based
on Cohen et al. \cite{cohen99}).}
\label{tab:calibrators}
\begin{tabular}{llrlll}
\hline\hline
Star & Sp. Type & $K$ & $\Theta_\mathrm{UD}$ & $\sigma(\Theta)$ & 
$T_\mathrm{eff}$ \\\hline
\object{56 Aql}         & K5 III    &  1.76 & 2.45 & 0.028 & 4046 \\
\object{31 Ori}         & K5 III    &  0.90 & 3.56 & 0.057 & 4046 \\
\object{58 Hya}         & K2.5 IIIb &  1.13 & 3.13 & 0.035 & 4318 \\
\object{66 Aql}         & K5 III    &  1.76 & 2.37 & 0.030 & 4046 \\
\object{70 Aql}         & K5 II     &  1.21 & 3.18 & 0.037 & 4064 \\
\object{$\theta$ Cen}   & K0- IIIb  & -0.26 & 5.32 & 0.058 & 4656 \\
\object{$\phi^1$ Aqr}   & K1- III   &  1.79 & 2.18 & 0.025 & 4508 \\
\object{$\lambda$ Gru}  & K3 III    &  1.44 & 2.64 & 0.030 & 4256 \\
\object{$\lambda$ Sgr}  & K1 IIIb   &  0.40 & 4.13 & 0.047 & 4508 \\
\object{$\pi^2$ Ori}    & K0 IIIb   &  1.69 & 2.14 & 0.023 & 4656 \\
\object{$\chi$ Phe}     & K5 III    &  1.52 & 2.69 & 0.032 & 4046 \\\hline
\end{tabular}
\end{table}

\subsection{VLTI/VINCI data reduction and calibration}
We computed mean coherence factors for each series of interferograms
using the VINCI data reduction software (version 3.0) by Kervella et al.
(\cite{kervella04}) employing the results based on the wavelets power
spectral density.
The calibration of the visibility values was performed as in Paper\,II
using a weighted average of the transfer function values obtained during 
the night.
\begin{table}
\centering
\caption{Details of our VLTI/VINCI observations of $\gamma$\,Sagittae
(date and time of observation, spatial frequency, azimuth angle of the 
projected baseline (E of N)), together with the measured squared 
visibility amplitudes and their errors. The last column denotes the 
number of successfully processed interferograms for each series. 
The effective wavelength for our $\gamma$\,Sagittae observations 
is $\sim$\,2.19\,$\mu$m. For each date of observation, we list the
equivalent uniform disc (UD) diameter obtained from only the data of
the specific night. Using all data together, we obtain an equivalent 
UD diameter of $\Theta_\mathrm{UD}=5.93\pm0.02$\,mas, or an equivalent
FDD diameter of $\Theta_\mathrm{FDD}=6.69\pm0.02$\,mas.}
\label{tab:vltiresults}
\begin{tabular}{lccccr}
\hline\hline
 UT & Sp. freq  & az  & $V^2$  & $\sigma_{V^2}$ & \# \\
 &[1/$^{\prime\prime}$]& [deg] &   &  &   \\\hline
\multicolumn{6}{c}{28 June 2002, $\Theta_\mathrm{UD}=5.91\pm0.03$\,mas}\\
05:16:57 & 131.14 & 136.63 & 1.825e-01 &  8.074e-03 &  161\\
05:22:40 & 130.08 & 136.65 & 1.828e-01 &  8.852e-03 &  152\\
05:32:34 & 128.15 & 136.73 & 2.077e-01 &  7.578e-03 &  172\\
06:44:22 & 111.69 & 139.42 & 3.112e-01 &  1.079e-02 &   80\\
\multicolumn{6}{c}{8 July 2002, $\Theta_\mathrm{UD}=5.89\pm0.04$\,mas}\\
05:12:10 & 124.21 & 137.05 & 2.241e-01 &  7.199e-03 &  397\\
05:20:13 & 122.44 & 137.27 & 2.426e-01 &  7.846e-03 &  418\\
\multicolumn{6}{c}{11 July 2002, $\Theta_\mathrm{UD}=5.98\pm0.04$\,mas}\\
03:06:45 & 142.16 & 138.28 & 1.064e-01 &  1.256e-02 &   55\\
05:29:49 & 117.47 & 138.08 & 2.495e-01 &  8.775e-03 &   69\\
05:34:22 & 116.38 & 138.31 & 2.660e-01 &  9.929e-03 &  207\\
05:40:41 & 114.83 & 138.64 & 2.878e-01 &  8.524e-03 &  297\\
05:47:01 & 113.25 & 139.02 & 2.977e-01 &  1.228e-02 &  198\\
\multicolumn{6}{c}{15 July 2002, $\Theta_\mathrm{UD}=5.94\pm0.05$\,mas}\\
04:22:40 & 128.76 & 136.70 & 1.926e-01 &  6.824e-03 &   90\\
04:41:27 & 124.90 & 136.98 & 2.223e-01 &  1.315e-02 &  139\\
\multicolumn{6}{c}{8 August 2002, $\Theta_\mathrm{UD}=5.92\pm0.03$\,mas}\\
03:05:49 & 125.17 & 136.95 & 2.179e-01 &  7.678e-03 &  469\\
03:12:20 & 123.77 & 137.10 & 2.244e-01 &  9.005e-03 &  436\\
03:18:34 & 122.39 & 137.27 & 2.309e-01 &  7.979e-03 &  468\\
03:24:34 & 121.03 & 137.47 & 2.471e-01 &  1.901e-02 &  178\\
04:13:56 & 108.94 & 140.21 & 3.376e-01 &  1.212e-02 &  423\\
\multicolumn{6}{c}{12 September 2002, $\Theta_\mathrm{UD}=5.99\pm0.11$\,mas}\\
01:26:22 & 116.44 & 138.29 & 2.607e-01 &  1.547e-02 &  190\\
01:47:17 & 111.24 & 139.54 & 3.407e-01 &  4.631e-02 &  105\\
\multicolumn{6}{c}{18 September 2002, $\Theta_\mathrm{UD}=5.88\pm0.05$\,mas}\\
00:32:53 & 123.38 & 137.15 & 2.347e-01 &  1.081e-02 &  428\\
00:40:53 & 121.59 & 137.39 & 2.393e-01 &  1.177e-02 &  407\\
00:47:14 & 120.14 & 137.61 & 2.667e-01 &  1.751e-02 &  210\\\hline
\end{tabular}
\end{table}
\subsection{VLTI/VINCI results}
Table~\ref{tab:vltiresults} shows the observational details together
with the resulting calibrated squared visibility amplitudes for each 
series of $\gamma$\,Sagittae interferograms. The listed errors include
the scatter of the coherence factors of the single scans, the errors of 
the adopted diameter values of the calibration stars, and the variation
of the obtained transfer function during each observing night.

As a first characterisation of the $K$-band stellar angular diameter,
we compute the equivalent UD ($I=\mu^0$) and FDD ($I=\mu^1$) diameters, 
as for our NPOI data. The broad-band squared visibility
amplitudes for the VINCI bandpass ($K$-band) are computed as
\begin{equation}
\label{eq:broadvinci}
|V_K|^2=\frac{\int_{0}^{\infty}\,F_\nu^2\,S_\nu^2\,|V(\nu)|^2\,d\nu}{\int_{0}^{\infty} F_\nu^2\,S_\nu^2\,d\nu},
\end{equation}
where $V(\nu)$ is the monochromatic visibility, $F_\nu$ is the stellar 
flux (assumed as Planck radiation) and 
$S_\nu$ the VINCI sensitivity function including the transmission of the
atmosphere, the optical fibers, the VINCI $K$-band filter, and the 
detector quantum efficiency. 

Note that in the case of VLTI/VINCI the {\it squared} visibility amplitudes
are integrated (Eq.~\ref{eq:broadvinci}), while in the case of 
NPOI the visibility amplitudes have first to be integrated and squared 
thereafter (Eq.~\ref{eq:broadnpoi}), see the note 
in Sect.~\ref{sec:npoiresults}.

Table~\ref{tab:vincifit} lists the obtained
diameter values for our VINCI data. 
\begin{table}
\centering
\caption{Fit results of our VINCI data to UD and FDD models.}
\label{tab:vincifit}
\begin{tabular}{llrr}
\hline\hline
Model & Diameter & $\chi^2_\nu$ \\\hline
UD    & $\Theta_\mathrm{UD}=5.93\pm 0.02$\,mas  & 0.63 \\
FDD   & $\Theta_\mathrm{FDD}=6.69\pm 0.02$\,mas  & 0.63 \\\hline
\end{tabular}
\end{table}
\begin{figure}
\resizebox{\hsize}{!}{\includegraphics{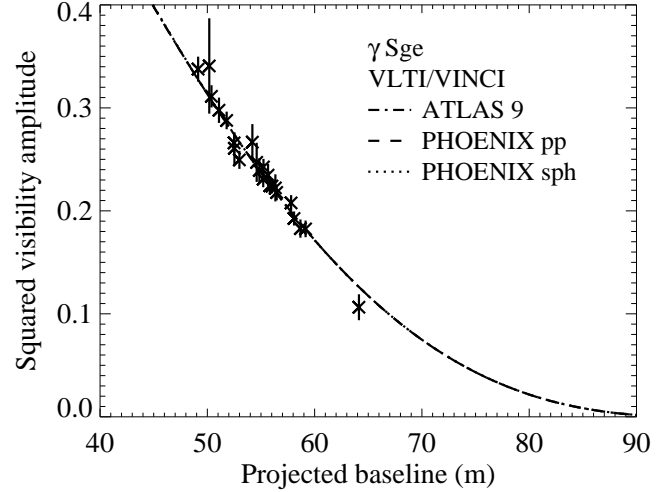}}
\caption{Measured $\gamma$\,Sagittae squared visibility amplitudes obtained
with VLTI/VINCI in June to September 2002, together
with the synthetic visibility curves of the best fitting models.}
\label{fig:vltivis}
\end{figure}
Figure~\ref{fig:vltivis} shows our obtained VINCI squared visibility 
amplitudes of $\gamma$\,Sge together with the best-fitting models
with parameters listed below in Table~\ref{tab:results}.
Since our VINCI data cover only one bandpass and only data of the first
lobe of the visibility function, it is -contrary to our NPOI data-
not feasible to constrain the limb-darkening effect solely based
on these VINCI data. This is also reflected by equal $\chi^2_\nu$ values 
obtained for UD and FDD models as well as by the virtually identical model 
visibility curves in Fig.~\ref{fig:vltivis}.

The increased equivalent UD diameter with respect to the 
shorter NPOI wavelengths is consistent with the general trend of 
decreasing strength of the limb-darkening effect with increasing wavelength. 
A detailed comparison of our data to model atmospheres follows below 
in Sect.~\ref{sec:models}.
\paragraph{Analysis of calibration uncertainties}
In order to test and verify the calibration uncertainties that are used
in our analysis, we investigate the night-to-night variation of the
obtained diameter values.
All derived single nights' diameter values and uncertainties are
consistent within 1.3$\sigma$ with the value obtained from 
all data together (5.93\,mas) as well as with the weighted mean 
of the single nights' values (5.92\,mas).

This confirms that our diameter value
is reliable and that our estimate of uncertainties 
is realistic. 
The obtained high-precision (0.3\%) UD and FDD diameter values 
of $\Theta_\mathrm{UD}=5.93\pm 0.02$\,mas 
and $\Theta_\mathrm{FDD}=6.69\pm 0.02$\,mas can thus be used without
further uncertainties.

Additional possible systematic errors that are constant over time scales 
larger than covered by our analysis, i.e. about 2 months, can not be 
ruled out. Such systematic errors could in principle be related to the 
calibration of the interferometric array and the instrument, such as the 
calibration  of the baseline length or the effective wavelength. Such 
uncertainties are not expected to represent a considerable source of error.
\section{Comparison to predictions by model atmospheres}
\label{sec:models}
\subsection{Employed model atmospheres}
We compare our measured visibility data to predictions by theoretical
model atmospheres in order to calibrate and test these models, and to 
derive fundamental stellar parameters of $\gamma$\,Sge.

We use plane-parallel {\tt ATLAS\,9}  (Kurucz \cite{kurucz93})
as well as plane-parallel and spherical {\tt PHOENIX} 
(Hauschildt et al. \cite{hauschildt99}) model atmospheres to calculate 
synthetic visibility data, as done in Papers I\&II.
We refer to the descriptions in Papers I\&II for more details on the 
employed model atmosphere files and their use.
Differences between {\tt ATLAS\,9} and {\tt PHOENIX} models include
different opacity tables, a different sampling of the model wavelengths,
a different sampling of the angles ($\mu$ values), 
and convective overshooting that is taken into account for {\tt ATLAS\,9}, 
but not for {\tt PHOENIX}.

The most important stellar input parameters for the 
plane-parallel models are effective 
temperature $T_\mathrm{eff}$ and surface gravity
$\log g$, and for the spherical {\tt PHOENIX} models in addition the 
mass $M$. We use solar chemical abundance, as 
appropriate for local cool giants. The values of $T_\mathrm{eff}$, $\log g$, 
and $M$ are already well constrained for $\gamma$\,Sagittae, as outlined 
in the Introduction (Sect.~\ref{sec:introduction}), namely 
$T_\mathrm{eff}\sim$\,3768\,K, $\log g\sim$\,1.06, $M\sim$\,1.3\,$M_\odot$.

The closest model of the {\tt ATLAS\,9} grid is the one for 
$T_\mathrm{eff}=$\,3750\,K and $\log g$\,=1.0 (see Papers I\&II for details
on the model file used). 
We have constructed a corresponding plane-parallel {\tt PHOENIX} model 
atmosphere with parameters $T_\mathrm{eff}=$\,3750\,K, $\log g$\,=1, as well
as a spherical {\tt PHOENIX} model atmosphere
with parameters $T_\mathrm{eff}=$\,3750\,K, $\log g$\,=1.0,
and $M=$\,1.3 (see Paper II for details on the model files).

\subsection{Calculation of synthetic visibility data}
We take into full account the bandpasses of our observations by integrating
the synthetic visibility data of monochromatic intensity profiles for 
each spectral channel of NPOI and for the $K$-bandpass of VLTI/VINCI. 
This ensures that the synthetic visibility values fully resemble the 
true bandpasses used for the observations and that they include the 
model-predicted effects from atomic lines and molecular bands for each
of our spectral channels. Monochromatic visibility values at frequency $\nu$
are calculated as (cf. Davis et al. \cite{davis00}, Eq.~6 from Paper I,
Eq.~1 from Paper II)
\begin{equation}
V_\mathrm{LD}(\nu)=\int_0^1\,(I_\nu(\mu)/I_\nu(0))\,J_0(\pi\,\Theta_\mathrm{LD}\,(B/\lambda))\,\mu\,d\mu.
\end{equation}
Here, $I_\nu(\mu)/I_\nu(0)$ is the normalised tabulated intensity profile,
which is an output of the model atmosphere. $J_0$ is the Bessel function
of first kind and order 0; $\Theta_\mathrm{LD}$ is the limb-darkened angular
diameter at which the intensity profile reaches 0; $B$ is the projected
baseline length.
Note that the evaluation of this integral is vulnerable
to numerical artifacts. We chose to use a linear interpolation of the
irregularly tabulated $I(\mu)$ model values onto a regular grid of 
1000 $\mu$ values between 0 and 1. The evaluation of the integral was then
performed using the Romberg method. Numerical results were checked against
analytical results for UD and FDD cases, and the resulting visibility
function for other cases was inspected for irregularities.

Broad-band visibility values integrated over the bandpasses of our
NPOI spectral channels and VINCI sensitivity function are calculated using 
Eqs.~\ref{eq:broadnpoi} and \ref{eq:broadvinci}, respectively.

\subsection{Calculation of best fitting angular diameters}
We calculate the best fitting limb-darkened (0\% intensity) 
angular diameter $\Theta_\mathrm{LD}$ as described above for each 
of these three model atmospheres (plane-parallel {\tt ATLAS\,9}, 
plane-parallel {\tt PHOENIX}, spherical {\tt PHOENIX} models) 
and for each of our two data sets (NPOI and VLTI/VINCI).
For each model fit, we treat $\Theta_\mathrm{LD}$ as the only
free parameter, and use all NPOI visibility data (squared visibility
amplitudes, triple amplitudes, and closure phases), a total of 644
data points, simultaneously. The fit is a standard least-square fit,
and optimises the total $\chi^2$ value of all 644 NPOI
data points.

As discussed in Sect.~3.4 of Paper\,II, models based on plane-parallel
geometry are optically thick from all viewing angles, the intensity
steeply dropping to 0 directly at the stellar limb. A plane-parallel 
model has, by definition, an atmosphere with an negligible thickness 
relative to the stellar radius.
Therefore, since any depth in such an atmosphere has a radius
effectively equal to the stellar radius, a Rosseland diameter
$\Theta_\mathrm{Ross}$ in such a geometry is equivalent to the
limb-darkened (or 0\% intensity) diameter $\Theta_\mathrm{LD}$.

Intensity profiles based on atmosphere models with spherical geometry, 
exhibit an inflection point and steepest decrease at radii smaller than 
the outermost model radius. The Rosseland mean optical depth increases
slowly for increasing angles $\mu$. 
Here, the ratio of the Rosseland diameter $\Theta_\mathrm{Ross}$ and
the 0\% intensity diameter $\Theta_\mathrm{LD}$ differs from unity, and 
this ratio $C_\mathrm{Ross/LD}$ is model-dependent and can be 
derived from the structure of the model atmosphere. This value depends
on the definition of the outermost radius $R_0$ of the model. $R_0$ of 
the spherical {\tt PHOENIX} model used here is given by the
standard boundary conditions, which are a continuum optical depth 
of 1e-6 at 1.2\,$\mu$m and an outer gas pressure of 1e-4\,dynes/cm$^2$
(see Paper\,II).
\begin{table*}
\centering
\caption{Results for the fit of {\tt ATLAS\,9} and {\tt PHOENIX} model
atmospheres to our interferometric VLTI/VINCI and NPOI data sets
of $\gamma$\,Sagittae.}
\label{tab:results}
\begin{tabular}{lrr}
\hline\hline
Model atmosphere & NPOI (526\,nm to 852\,nm)& VLTI/VINCI (2190\,nm)\\\hline
{\tt ATLAS\,9}, plane-parallel, $T_\mathrm{eff}=$\,3750\,K, $\log g$\,=1.0 & 
$\Theta_\mathrm{LD}=6.18\pm0.06$\,mas & $\Theta_\mathrm{LD}=6.08\pm0.02$\,mas\\
& $\chi^2_\nu=2.2$  & $\chi^2_\nu=0.6$  \\[4ex]
{\tt PHOENIX}, plane-parallel, $T_\mathrm{eff}=3750$\,K, $\log g=1.0$ &
$\Theta_\mathrm{LD}=6.11\pm0.06$\,mas & $\Theta_\mathrm{LD}=6.09\pm0.02$\,mas\\
& $\chi^2_\nu=2.3$ & $\chi^2_\nu=0.6$ \\[4ex]
{\tt PHOENIX}, spherical, $T_\mathrm{eff}=3750$\,K, $\log g=1.0$, $M=1.3\,M_\odot$ &
$\Theta_\mathrm{LD}=6.30\pm0.06$\,mas & $\Theta_\mathrm{LD}=6.34\pm0.02$\,mas\\
& $\chi^2_\nu=2.4$ & $\chi^2_\nu=0.6$ \\
&$\Theta_\mathrm{Ross}=6.02\pm0.06$\,mas& $\Theta_\mathrm{Ross}=6.06\pm0.02$\,mas\\
& \multicolumn{2}{c}{$\overline{\Theta_\mathrm{Ross}}=6.06\pm0.02$\,mas}\\\hline 
\end{tabular}
\end{table*}
\subsection{Results and discussion}
\paragraph{Results} Table~\ref{tab:results} shows, separately
for our NPOI and VINCI data sets, the resulting best-fitting angular
diameter values based on the different considered model atmospheres,
together with the corresponding $\chi^2_\nu$ values.
For the spherical model atmospheres, the 0\% intensity diameter
$\Theta_\mathrm{LD}$ is transformed to the Rosseland diameter
$\Theta_\mathrm{Ross}$ as described above. 

The corresponding synthetic visibility data are compared
to the measured data in Figs.~\ref{fig:npoivis1} to \ref{fig:npoiphas} 
for our NPOI data set and in Fig.~\ref{fig:vltivis} for our VLTI/VINCI 
data set. 

\paragraph{Best-fitting angular diameters}
Based on the plane-parallel {\tt ATLAS 9} model and our NPOI data, 
we reproduce the limb-darkened diameter 
$\Theta_\mathrm{LD}=6.18\pm0.06$\,mas from Paper\,I, despite the greater
usable wavelength range and higher precision of the current NPOI data. The
error includes an adopted 1\% systematic error due to the NPOI 
wavelength calibration (the formal error is 0.004\,mas). The VLTI/VINCI
diameter for this model atmosphere of $\Theta_\mathrm{LD}=6.08\pm0.02$\,mas
is not well consistent with the NPOI diameter 
($\approx$\,2\,$\sigma$ difference).

The plane-parallel {\tt PHOENIX} model leaves the near-infrared
VLTI/VINCI diameter quasi unchanged (6.09 mas compared to 6.08 mas)
with respect to the plane-parallel {\tt
ATLAS} model, while it results in a smaller (by $\approx$\,1\,$\sigma$)
visual NPOI diameter compared to the {\tt ATLAS} model.  A comparison
of these models' temperature structures reveals that the {\tt ATLAS}
model exhibits a steeper temperature gradient relative to the
plane-parallel {\tt PHOENIX} near Rosseland optical depth unity.  This
steeper gradient leads to stronger limb darkening at NPOI wavelengths
and consequently a larger best angular diameter.  The shallower
temperature gradient of the plane-parallel {\tt PHOENIX} model leads
to a better agreement between the NPOI and VLTI/VINCI diameters.

Finally, the spherical {\tt PHOENIX} model leads to a Rosseland
angular diameter of $\Theta_\mathrm{Ross}=6.02\pm 0.06$\,mas for the NPOI
data set and $\Theta_\mathrm{Ross}=6.06\pm 0.02$\,mas for the VLTI/VINCI 
data set. The larger best-fit diameters for the plane-parallel {\tt PHOENIX} 
model compared to the spherical {\tt PHOENIX} model appears to be
due to model geometry.  The agreement of NPOI and VLTI/VINCI data sets
within their 1\,$\sigma$ error bars gives confidence in both, the atmosphere 
models and the accuracy of the results from NPOI and VLTI/VINCI. 
The weighted mean of the NPOI and VLTI/VINCI results is  
$\Theta_\mathrm{Ross}=6.06\pm 0.02$\,mas.

\paragraph{Shape of the visibility function}
The measured and model-predicted visibility functions are generally
consistent.
However, the obtained reduced $\chi^2_\nu$ values for the NPOI data
between 2.2 and 2.4 are above unity, as would be expected 
for a perfect match. This indicates differences at the 2$\sigma$ level
between observed visibility data and the model predictions.

These differences are most evident in 
(1) a lower second maximum of the 
measured visibility function with respect to the model prediction
on the EW baseline (Fig.~\ref{fig:npoivis1}), and (2) a flattened measured
visibility function with respect to the model predictions
at the blue end on the CW baseline 
(Fig.~\ref{fig:npoivis2}).
It is not yet clear if and by how far these deviations of 
measured and synthetic visibility functions indicate 
different details of the limb-darkening effect at visual spectral channels,
or if they are caused by additional calibration uncertainties that are
not included in the error bars.
In particular the flattening of the measured visibility function at the
bluest spectral channels on the CW baseline can most likely be explained
by additional calibration uncertainties of our NPOI data, as the 
instrumental transfer function for these data exhibits a drop which
may not be fully compensated.

The obtained diameter values in Table~\ref{tab:results} are not affected
by possible calibration uncertainties since the best-fitting 
diameter for any given model atmosphere is mostly constrained by the position 
of the first minimum and the global shape of the visibility curve.

At optical wavelengths including all our NPOI spectral channels, 
TiO absorption bands are very important for the modelling of atmospheres
of cool giants. It has been shown that the use of different line list 
combinations of TiO and H$_2$O leads to significantly different model 
structures and spectra, in particular in the optical where TiO bands are
important (Allard et al. \cite{allard00}). A possible explanation 
for differences between our visibility data and the model predictions 
could thus also be mismatching opacity tables for the TiO bands and/or a 
mismatching spatial structure of the layers where TiO molecules reside.

In order to estimate the effect of a lower model strength of the limb-darkening
effect on the obtained diameter value, we used a spherical {\tt PHOENIX} model 
with $T_\mathrm{eff}=3500$\,K instead of our favourite model with
$T_\mathrm{eff}=3750$\,K (other parameters unchanged). The height of the 
second maximum of the visibility function at a wavelength of 600\,nm
is reduced from $\approx\ 0.0085$ (see the lower right panel of Fig.~\ref{fig:npoivis1})
to $\approx\ 0.0070$. The obtained best-fitting diameter value for the NPOI data
set changes
from $\Theta_\mathrm{Ross}=6.02\ \pm\ 0.06$\,mas to 
$\Theta_\mathrm{Ross}=6.03\ \pm\ 0.06$\,mas, and is unchanged for 
the VLTI/VINCI data set. This shows that an imperfect modelling of the
strength of the limb-darkening effect within our uncertainties
does not have a significant effect on our obtained diameter values for $\gamma$\,Sge.
\paragraph{Model atmosphere fluxes}
\begin{figure}
\centering
\resizebox{\hsize}{!}{\includegraphics{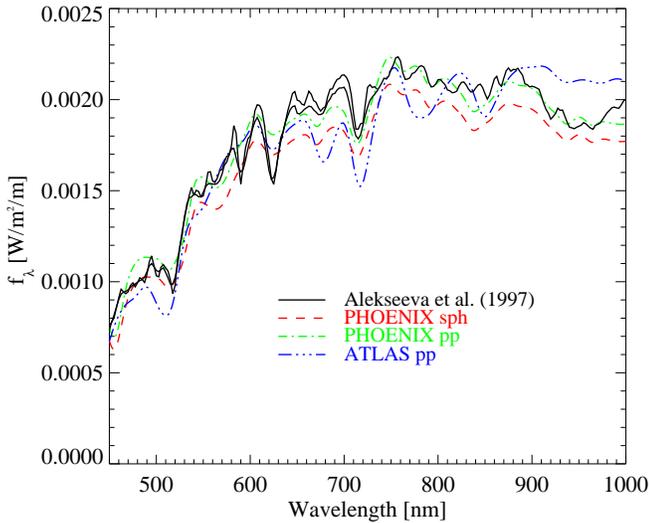}}
\caption{Flux of $\gamma$\,Sge from Alekseeva et al. (\cite{alekseeva97})
in the wavelength range of our NPOI observations, compared to the 
model atmosphere predictions with model parameters listed 
in Table~\ref{tab:results}. The 
limb-darkened 0\% diameter values $\Theta_\mathrm{LD}$ derived
from the fit to the interferometric data were used to scale the model SEDs.
The spectral resolution of the model SEDs is convolved to the 
resolution of the data used, i.e. to 10\,nm}.
\label{fig:fluxes}
\end{figure}
Figure~\ref{fig:fluxes} shows the measured flux of $\gamma$\,Sge
from Alekseeva et al. \cite{alekseeva97} in the wavelength range
from 0.4-1.0\,$\mu$m, i.e. covering the NPOI range used in this paper.
Also shown are the predictions by the model atmospheres with
parameters listed in Table~\ref{tab:results}. 
The limb-darkened 0\% diameter values $\Theta_\mathrm{LD}$ derived
from the fit to the interferometric data were used to scale the model SEDs.
The spectral resolution of the model SEDs is convolved to the 
resolution of the data used, i.e. to 10\,nm.
The model-predicted
flux curves based on the three considered models are well consistent with 
the general shape of the measured flux, while the detailed description
of the spectral bands and features differs between the different models
and the measured values.  
These differences can most likely be explained by the treatment of TiO 
absorption lines which are important for the visual wavelength range
and difficult to model (Allard et al. \cite{allard00}), as mentioned 
in the paragraph above.
\paragraph{Deviations from circular symmetry}
Differences between data and models are also observed for the closure phases
(Fig.~\ref{fig:npoiphas}). The observed smooth variation of the 
closure phases from
0 to $\pi$ instead of the expected instantaneous flip may indicate
a small deviation from spherical symmetry as already mentioned in Paper~I.
As our maximum spatial resolution reaches 2.7\,mas, and the stellar disc
has a size of $\Theta_\mathrm{Ross}=6.06$\,mas, the stellar disc is well
resolved with 2.2 resolution elements across the disc. As a result,
our data is sensitive to small deviations of circular symmetry.
Such a deviation can in principle be caused by surface features such as spots, 
an asymmetric shape of the photosphere or of more extended (molecular) 
layers, or a faint unknown companion. 
Our target $\gamma$\,Sge shows a relatively high 
photospheric pressure scale height 
of $H_\mathrm{P0}=R_\mathrm{gas} T_\mathrm{eff}/g\approx 0.006$\,R$_\star$.
Relatively large-scale ($\approx$\,0.06\,R$_\star$) surface 
inhomogeneities caused by convection could thus be expected 
(cf. Freytag et al. \cite{freytag97}).

\paragraph{Spherical versus plane-parallel model geometry}
The synthetic visibility data based on the same {\tt PHOENIX} models that
solely differ by spherical versus plane-parallel model geometry are
virtually identical (Figs.~\ref{fig:npoivis1}-\ref{fig:npoiphas}).
Thus, the model geometry has no noticeable effect on the shape of the 
visibility and can not be constrained by our visibility data.
However, the derived angular diameter values differ by 1.5\% for the 
visual NPOI data and by 0.5\% for the near-infrared VLTI/VINCI data. The
spherical geometry allows us to more precisely define the 
stellar Rosseland radius with respect to the outermost model layer, and is
thus more reliable than the angular diameter obtained from the plane-parallel
model. 
As already noticed in Paper\,II for the M4 giant $\psi$\,Phe, 
the 0\% intensity (LD) diameter based on a plane-parallel model seems 
to somewhat overestimate the stellar diameter with respect to the 
Rosseland diameter based on a spherical model. Our present results indicate
a wavelength-dependent amount of this overestimation.
\section{Summary and conclusions}
\label{sec:discussion}
We have compared our visual $\gamma$\,Sagittae NPOI visibility data 
for 19 spectral channels with central wavelengths between 526\,nm to 852\,nm
as well as our near-infrared $K$-band VLTI/VINCI  visibility data
with effective wavelength 2.19$\mu$m to a
plane-parallel {\tt ATLAS\,9}, a plane-parallel {\tt PHOENIX}, and
a spherical {\tt PHOENIX} model atmosphere. The stellar parameters
effective temperature $T_\mathrm{eff}$, surface gravity $\log g$,
and $M$ of the model atmospheres used were fixed a-priori based on
previous information on this star.

The spherical geometry of the {\tt PHOENIX} model enables us to
precisely define the Rosseland radius of the star with respect to the
outermost model layer and thus the 0\% intensity diameter. 
This model leads to consistent Rosseland angular diameters for 
our NPOI and VLTI/VINCI data sets. This agreement increases the confidence 
in the model atmosphere predictions from optical to near-infrared wavelengths
as well as in the calibration and accuracy of both interferometric facilities.
In addition, the consistent angular diameter derived from our
VLTI/VINCI data on a night-by-night basis over a total range of about 2
months increases confidence in the given calibration uncertainties.

\begin{table}
\centering 
\caption{Revised fundamental parameters of the M0 giant $\gamma$\,Sagittae
based on the analysis of this paper. For the details of the calculation,
see the text.}
\label{tab:fundpar}
\begin{tabular}{ll}
\hline\hline
Parameter & Value \\\hline
Rosseland angular diameter & $\Theta_\mathrm{Ross}=6.06\pm 0.02$\,mas\\
Rosseland linear radius  & $R_\mathrm{Ross}=55\pm 4 R_\odot$\\
Bolometric flux & $f_\mathrm{bol}=(2.57\pm 0.13)\times 10^{-9}$\,W/m$^2$\\
Effective temperature & $T_\mathrm{eff}=3805\pm 55$\,K\\
Luminosity & $\log L/L_\odot=2.75\pm 0.08$\\
Mass & $M=1.4\pm 0.4 M_\odot$\\
Surface gravity & $\log g=1.1\pm 0.2$\\\hline
\end{tabular}
\end{table}
The Rosseland angular diameter of $\gamma$\,Sagittae of
$\Theta_\mathrm{Ross}=6.06\pm 0.02$\,mas, based on the comparison
of our NPOI and VLTI/VINCI data to the spherical {\tt PHOENIX}
model, corresponds to a Rosseland linear radius 
of $R_\mathrm{Ross}=55\pm 4 R_\odot$, derived with the 
Hipparcos parallax of $\pi=11.90\pm 0.71$\,mas. The error of the
Rosseland linear radius is dominated by the uncertainty of the parallax,
not by the precision of our interferometric measurement.
With the  bolometric 
flux $f_\mathrm{bol}=(2.57\pm 0.13)\times 10^{-9}$\,W/m$^2$ 
(Sect.~\ref{sec:introduction}) and the Rosseland angular diameter, 
the effective temperature is constrained to
$T_\mathrm{eff}=3805\pm 55$\,K. Again, the major contribution to this
error originates from the uncertainty in $f_\mathrm{bol}$ and not
from our interferometric measurement.
The Rosseland linear radius and $T_\mathrm{eff}$ result in a 
luminosity of $\log L/L_\odot=2.75\pm 0.08$.
Placing $\gamma$\,Sagittae
on the Hertzsprung Russel diagram using these values, and comparing to
stellar evolutionary tracks by Girardi et al. (\cite{girardi00}) as
in Paper\,II (Fig.~1 of Paper\,II) we can estimate a mass 
of $M=1.4\pm 0.4 M_\odot$, 
and thus a surface gravity of $\log g=1.1\pm 0.2$.
Table~\ref{tab:fundpar} summarises our revised values of $\gamma$\,Sagittae's
fundamental parameters.

The closure phases show a smooth transition from 0 to $\pi$ rather than 
a sharp flip, which could be due to a small deviation from 
circular symmetry of the well resolved stellar disc due to surface 
features such as spots, an asymmetric extended molecular layer, 
or a faint companion.
\begin{acknowledgements}
This work was performed in part under contract with the Jet Propulsion
Laboratory (JPL) funded by NASA through a Michelson Fellowship Program (JPA).
\end{acknowledgements}
\end{document}